\documentclass[
 aip, jcp,
 amsmath,amssymb,
 reprint,
]{revtex4-1}

\usepackage[ruled,vlined,linesnumbered]{algorithm2e}
\usepackage[most]{tcolorbox}

\newtcolorbox{algobox}{
  enhanced,
  colback=white,
  colframe=black!55,
  boxrule=0.6pt,
  arc=1.5pt,
  left=6pt,
  right=6pt,
  top=5pt,
  bottom=5pt,
  width=0.88\linewidth,
  center
}

\usepackage{graphicx}
\usepackage{dcolumn}
\usepackage{bm}

\usepackage[utf8]{inputenc}
\usepackage[T1]{fontenc}
\usepackage{mathptmx}
\usepackage{etoolbox}
 \usepackage{booktabs}
\usepackage{lmodern}
\usepackage{siunitx}
\usepackage[table]{xcolor}
\usepackage{makecell}
\definecolor{hlbase}{HTML}{3A53C8}
\colorlet{firsthl}{hlbase!32}
\colorlet{secondhl}{hlbase!15}
\newcommand{\dev}[1]{\textsubscript{\textcolor{gray}{#1}}}
\usepackage[colorlinks,allcolors=black,citecolor=blue,urlcolor=blue,breaklinks]{hyperref}
\usepackage[mathlines]{lineno}

\newcommand{\figref}[2]{Figure~\ref{#1}#2}

\definecolor{algcomment}{HTML}{6B7280}

\SetCommentSty{algcommentfont}
\SetKwComment{Comment}{$\triangleright$\ }{}

\makeatletter
\patchcmd{\@makecaption}{\centering}{\raggedright}{}{}
\renewcommand{\figurename}{Figure}
\renewcommand{\fnum@figure}{\sffamily{\bfseries\figurename~\thefigure}}
\renewcommand{\fnum@table}{\sffamily\tablename~\thetable}
\patchcmd{\@makecaption}{\small}{\small\sffamily}{}{}
\makeatother

\makeatletter
\def\@email#1#2{
 \endgroup
 \patchcmd{\titleblock@produce}
  {\frontmatter@RRAPformat}
  {\frontmatter@RRAPformat{\produce@RRAP{*#1\href{mailto:#2}{#2}}}\frontmatter@RRAPformat}
  {}{}
}
\makeatother

\begin{document}

\preprint{AIP/123-QED}

\title[Rem3Di: Smooth, chiral 3D molecular descriptors]{Rem3Di: Learning smooth, chiral 3D molecular descriptors from atomistic foundation models}
\author{Steffen Wedig}
 \affiliation{
Cavendish Laboratory, Department of Physics, University of Cambridge, Cambridge CB3 0HE, U.K.
}
\affiliation{
Max Planck Institute for Polymer Research, Ackermannweg 10, Mainz, 55128, Germany
}
\author{Felix Burton}
\affiliation{
Cavendish Laboratory, Department of Physics, University of Cambridge, Cambridge CB3 0HE, U.K.
}
\author{Rokas Elijošius\textsuperscript{*}}
\affiliation{
Cavendish Laboratory, Department of Physics, University of Cambridge, Cambridge CB3 0HE, U.K.
}
\author{Christoph Schran\textsuperscript{*}}
\affiliation{
Cavendish Laboratory, Department of Physics, University of Cambridge, Cambridge CB3 0HE, U.K.
}

\author{Lars L. Schaaf\textsuperscript{*}}
\affiliation{
Cavendish Laboratory, Department of Physics, University of Cambridge, Cambridge CB3 0HE, U.K.
}
\affiliation{
Department of Materials, Imperial College London, Exhibition Road, London SW7 2AZ, U.K.
}
\email[Correspondence to: ]{re344@cam.ac.uk, cs2121@cam.ac.uk, lls34@cam.ac.uk}

\date{\today}

\begin{abstract}
Foundation machine-learned interatomic potentials (MLIPs) are trained on large quantum-mechanical datasets and generalise across broad regions of chemical and configurational space.
Beyond their usual role in accelerating sampling-based simulations, their internal representations encode chemically rich local atomic environments.
Here, we introduce Rem3Di, a representation-learning framework that repurposes latent features from atomistic foundation models as transferable molecular descriptors for property prediction and virtual screening.
Rem3Di combines a potential's per-atom features into a single fixed-length descriptor of the whole molecule that varies smoothly with three-dimensional structure and is invariant to the ordering of the atoms. The descriptor can be used directly or fine-tuned for specific prediction tasks.
To capture molecular handedness, Rem3Di constructs pseudoscalar features, which are unchanged by rotation but reverse sign under mirror reflection. This lets the descriptor distinguish enantiomers, which can differ in activity and toxicity.
The transformer is pretrained on large molecular datasets by reconstructing corrupted atom features, so no experimental labels are required.
Across public drug-property benchmarks, Rem3Di matches or exceeds published baselines without relying on classical 2D fingerprints.
Additionally, the same descriptor yields chemically meaningful differentiation of transition-metal complexes without predefined bonding rules or handcrafted representations. 
Rem3Di therefore provides a route from simulation-trained atomistic representations to transferable, chirality-aware molecular representations for chemical machine learning.
\end{abstract}

\maketitle

\begin{figure*}[tbh]
    \centering
    \includegraphics[width=0.9287988\linewidth]{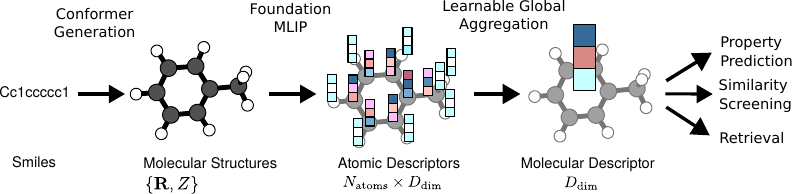}
    \caption{\textbf{Constructing 3D molecular descriptors with the Rem3Di framework.} Atom-centred features from atomistic foundation models are combined into a fixed-dimension molecular descriptor that varies smoothly with 3D structure.
    }
    \vspace{-.1cm}
    
    \label{fig:overview}
\end{figure*}

\section{Introduction}

The predictive power of molecular machine learning is often determined as much by the molecular representation as by the learning algorithm itself. A useful representation must be efficient to compute, compatible with statistical learning, and expressive enough to encode the structural features that determine molecular properties. Many successful workflows in drug discovery rely on two-dimensional graph-based descriptors, particularly Extended Connectivity Fingerprints (ECFPs), which encode atom environments through iterative hashing of molecular graphs~\cite{Morgan1965,Rogers2010}. These fingerprints are fast, robust, and remain competitive baselines when combined with simple supervised models such as random forests~\cite{Breiman2001,Jiang2021}. However, graph-based descriptors make bonding topology the primary object of representation. This limits their ability to describe conformational effects, stereochemistry, and systems for which bonding assignments are ambiguous, such as transition-metal complexes~\cite{Rasmussen2025}.

In parallel, a large body of work in atomistic machine learning has developed representations of chemical systems in three dimensions. These methods describe molecules and materials through atomic identities and spatial arrangements of the molecule's nuclei, rather than relying solely on an assigned network of molecular bonds. Global descriptors such as the Coulomb matrix and related fixed-size representations encode molecular structure from nuclear charges and coordinates~\cite{Rupp2012,Hansen2015,Huo2022}. Local atom-centred descriptors, including atom-centred symmetry functions~\cite{Behler2007, Behler2011}, SOAP-type atom-density representations~\cite{Bartok2013}, the Atomic Cluster Expansion~\cite{Drautz2019} and SLATM~\cite{Huang2020}, describe atomic environments through invariant geometric features. Molecular kernels based on local environments further showed how atom-level representations can be compared or aggregated into molecular similarities~\cite{De2016}. Many of these representations were developed for interatomic potentials, but related geometric descriptors and neural architectures have also been used directly for molecular property prediction~\cite{Schuett2017,Batatia2022,Batzner2022,Wang2024,Staerk2021,Zhou2022}. They established principles that remain central to atomistic learning, including locality, smoothness, completeness and rotational invariance~\cite{Pozdnyakov2020}.

Recent machine-learned interatomic potentials extend this line of work to learning atomic representations for predicting energies and forces from quantum mechanical calculations~\cite{Schuett2017, Batzner2022, Batatia2022}. Their internal features are therefore shaped by local chemical environments, energetic smoothness, and the symmetries of three-dimensional space. Large-scale models trained across broad chemical domains now generalise well beyond individual datasets and can be viewed as atomistic foundation models~\cite{Bommasani2021,Merchant2023,Batatia2024,Kovacs2025,Wood2025,Ple2025,Neumann2024,Rhodes2025,Fu2025}. This generalisability is not guaranteed for atom-scale models. Protein-ligand cofolding models, for example, show limited chemical generalisation, indicating that they may not learn transferable representations of chemical interactions~\cite{Masters2025,Abramson2024,Krishna2024,Boitreaud2024,Wohlwend2024}. The central question is therefore whether the latent features of atomistic foundation models can be reused directly as molecular descriptors.

Machine-learned interatomic potentials are most commonly used as fast surrogate energy models for sampling-based simulations, such as molecular dynamics. Although this has transformed atomistic simulation, many molecular observables still require extensive sampling and free-energy estimation, making such workflows difficult to deploy across large chemical libraries~\cite{Cournia2017}. An alternative is to bypass simulation and use the internal representations of MLIPs directly as molecular descriptors. This route is attractive because these features are learned from quantum-mechanical energies and forces, and therefore encode transferable information about local chemical environments. 
Recent studies have also analysed the structure and utility of atomistic foundation-model latent spaces, suggesting that independently trained models may converge towards shared representations of atomic environments~\cite{Huh2024,Li2026,Edamadaka2025,Uchiyama2026}. These findings support the idea that MLIP features contain transferable chemical information beyond their original use in force and energy prediction. Related work has begun to reuse atom-centred features from machine-learned potentials, either pooled into structure-level descriptors for property prediction~\cite{Kim2026} or used directly for atom-level properties, such as NMR chemical shifts in molecules~\cite{Shiota2024} and proteins~\cite{Bojan2025}.
However, converting atom-centred MLIP features into a useful molecular descriptor is non-trivial. The representation must aggregate a variable number of atoms into a fixed-size embedding, preserve smooth three-dimensional chemical information, remain invariant to rotations and atom permutations, and still distinguish stereoisomers. It must also be trainable in the low-label regimes typical of molecular property prediction.

Here, we introduce Rem3Di, a representation-learning framework that addresses these requirements (\figref{fig:overview}{}). Given a molecular structure, a frozen atomistic foundation model first produces atom-centred latent features. Rem3Di then learns to contract these local features into a fixed-dimensional, conformation-aware molecular embedding using permutation-invariant attention-based set pooling~\cite{Zaheer2017,Lee2018}. To make the descriptor sensitive to stereochemistry, Rem3Di constructs pseudoscalar channels from equivariant features. These channels are invariant to rotations but change sign under reflection, allowing the representation to distinguish enantiomers while respecting the symmetries of three-dimensional space. Finally, because the aggregation weights themselves must be learned, we introduce a self-supervised denoising objective that pretrains the molecular descriptor on large unlabelled molecular datasets. The model learns to reconstruct clean atom-level representations from corrupted MLIP features, enabling it to acquire chemically meaningful aggregation rules before task-specific labels are introduced.

We find that the chiral encoding generalises across molecular scaffolds and predicts the sign of the optical rotation of chiral molecules. Across public drug-property benchmarks, Rem3Di matches or improves upon descriptor-based and learned baselines on several regression tasks without relying on classical two-dimensional fingerprints. We further show that the same descriptor gives chemically meaningful differentiation of transition-metal complexes without predefined bonding rules. Rem3Di therefore provides a route from simulation-trained atomistic foundation models to transferable, chirality-aware molecular descriptors for chemical machine learning.

\begin{figure*}[tb]
    \centering
    \includegraphics[width=0.98599\linewidth]{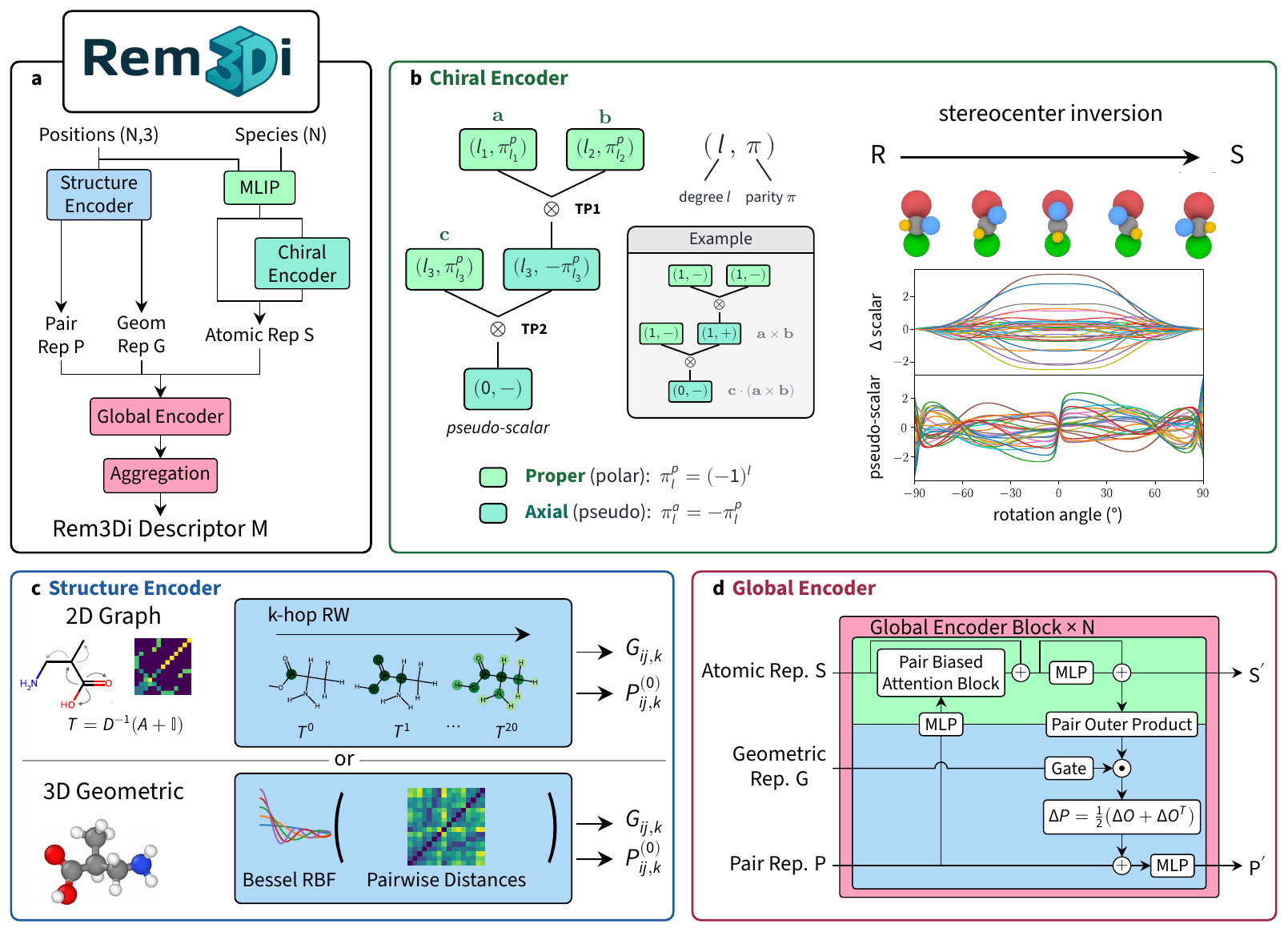}
    \caption{\textbf{The Rem3Di architecture.} (a) Overview of the Rem3Di blocks mapping position and chemical element to a molecular descriptor. (b-left) Construction of chiral features using two tensor products, with an example for vectorial features ($\ell_1=\ell_2=\ell_3=1$) corresponding to the vector triple product. Tensors $\mathbf a, \mathbf b$ and $\mathbf c$ are learnable combinations of latent features in a spherical harmonic basis. (b-right) Chiral atom-level embeddings of a molecule as it transforms smoothly into its mirror image through the continuous rotation of a functional group, showing the even and odd behaviour of the scalar and pseudoscalar features. (c) Two options for the structure encoder. (d) Architecture of the global encoder mixing atomic and pair representations.}
    \label{fig:method_figure}
\end{figure*}

\section{Method: The Rem3Di architecture}

\subsection{Overview}
Rem3Di (REpresentation learning for Molecules using 3D Information) maps a molecular conformer to a fixed-length molecular descriptor for property prediction, virtual screening, and stereochemical classification, as summarised in \figref{fig:overview}{}. The dimension of the descriptor is fixed at a given channel dimension and is independent of the number of atoms. The input is a set of atomic species and Cartesian coordinates, obtained either from an existing three-dimensional structure or generated from a molecular representation such as a SMILES string using standard conformer-generation tools such as RDKit. The architecture, shown in \figref{fig:method_figure}{a}, has four main components. A frozen atomistic foundation model first converts the conformer into atom-centred latent features. A chiral encoder then constructs pseudoscalar channels from equivariant MLIP features, making the atom representation sensitive to molecular handedness (\figref{fig:method_figure}{b}). Following architectures such as AlphaFold~\cite{jumper, Abramson2024}, we construct a pair representation that is updated throughout the architecture. A structure encoder builds an initial pair representation from the molecular 3D geometry (\figref{fig:method_figure}{c}). Finally, a global encoder mixes atom and pair information across the whole molecule, before permutation-invariant attention pooling produces the Rem3Di molecular descriptor (\figref{fig:method_figure}{d}). Detailed architectural hyperparameters are reported in Appendix~\ref{app:hyperparameters}.

Concretely, the frozen MLIP provides per-atom invariant features and, for equivariant backbones, higher-order tensor features. The invariant features are normalised and projected to the encoder width, while the equivariant features are passed to the chiral encoder to produce parity-odd pseudoscalars. These scalar and pseudoscalar channels form the initial atom representation $S^{0}$. In parallel, the structure encoder initialises a pair representation $P$ from interatomic distances and optional molecular graph information. The global encoder then updates $S$ and $P$ jointly using pair-biased self-attention, allowing local MLIP-derived atom features to be contextualised by the full molecular geometry. The final atom representations are aggregated by pooling-by-multihead-attention~\cite{Lee2018} into a fixed-dimensional descriptor, which can be used directly, adapted by low-rank fine-tuning, or trained further with the supervised task head.

\subsection{Foundation-model featurisation}
\label{sec:featurisation}

For each conformer, the pretrained MLIP is evaluated on the atomic species and coordinates, and internal per-atom node features are extracted. This is the entry stage of the pipeline in \figref{fig:method_figure}{a}. The framework is agnostic to the choice of backbone MLIPs and we show performance on equivariant (MACE) models and models where equivariance is learnt through data augmentation (Orb)~\cite{Rhodes2025}. We mainly focus on constructing meaningful latent spaces from equivariant architectures, so the chiral encoder, for example, is only relevant if the latent features are equivariant and can be separated by their spherical degree ($\ell$). In all experiments the MLIP is frozen, and no gradients are propagated through the potential. Rem3Di therefore learns a map from pretrained atom-level features to molecular descriptors, rather than retraining or fine-tuning the atomistic foundation model itself.

For equivariant MLIPs, the extracted features are organised as irreducible representations of $O(3)$. They include even scalar channels, which are invariant under rotations and reflections, and higher-order tensor channels, rotating with the atomic environment and acquiring a parity sign under reflection. The scalar channels are normalised using training-set statistics and projected to the encoder width. Given a fixed encoder width, the number of Rem3Di model parameters is hence predominantly independent of the MLIP's feature dimension. The equivariant channels are rescaled with an $O(3)$-equivariant RMS normalisation, which changes channel magnitudes but not tensor directions. These normalised equivariant features are passed to the chiral encoder. The standardised scalar features and the pseudoscalar output of the chiral encoder together form the initial atom representation $S^{0}\in\mathbb{R}^{N\times d_\text{model}}$. Backbone-specific feature choices, layer selection, and projection dimensions are given in Appendix~\ref{app:hyperparameters}.

\subsection{Constructing pseudoscalars from equivariant features}
\label{sec:app-pseudo-scalar}

The scalar latent features of common equivariant MLIPs are invariant under inversion ($\mathbf r\mapsto-\mathbf r$), and therefore cannot distinguish between two enantiomers~\cite{Dumitrescu2024}. This is appropriate for potential-energy prediction, since the potential energy is unchanged by reflection, but is insufficient for molecular properties that depend on interactions with chiral environments, such as proteins, enzymes, chromatographic media, or solvents. Rem3Di therefore constructs atom-wise pseudoscalar features from equivariant MLIP channels. A pseudoscalar is invariant under proper rotations but changes sign under reflection, making it the natural scalar representation of molecular handedness~\cite{Harris1999}. An alternative route to chirality-sensitive features takes determinants over ordered atomic quadruples around each stereocentre, which are likewise invariant under proper rotations and change sign under reflection~\cite{Shi2026}. As illustrated in \figref{fig:method_figure}{b}, reflection-even scalars are identical for mirror configurations, whereas pseudoscalars have equal magnitude and opposite sign. Along the continuous path connecting the two enantiomers, Rem3Di's pseudoscalars pass through zero when the atoms become coplanar at $\theta=0$, corresponding to an achiral configuration. Below we outline how the chiral encoder constructs pseudoscalars: we first fix the notation and parity conventions, then recall the spherical tensor product, and then show that two successive products are required to reach a pseudoscalar.

\subsubsection{Notation and parity}

The irreducible representations of $SO(3)$ are labelled by an angular degree $\ell$, and act on a $(2\ell+1)$-dimensional space whose elements transform under proper rotations $R$ through the Wigner $D$-matrices $D^{(\ell)}(R)$,
\begin{equation}
\left[R\cdot\mathbf t^{(\ell,\pi)}\right]_m
=\sum_{m'=-\ell}^{\ell}
D_{mm'}^{(\ell)}(R)\,
t_{m'}^{(\ell,\pi)},
\label{eq:wigner-d}
\end{equation}
where $D_{mm'}^{(\ell)}(R)$ is the matrix of the degree-$\ell$ irreducible representation in the spherical-harmonic basis. The components of spherical tensors, hence, transform like the spherical harmonics $Y_\ell^m$ and only components indexed by $m$ mix. Physical tensors also carry a behaviour under inversion $I:\mathbf r\mapsto-\mathbf r$, which is not fixed by $\ell$. We record it with a second index, the parity $\pi\in\{+1,-1\}$, defined by
\begin{equation}
I\cdot\mathbf t^{(\ell,\pi)}
=\pi\,\mathbf t^{(\ell,\pi)}.
\end{equation}
Adjoining inversion to the proper rotations extends the group from $SO(3)$ to the full orthogonal group $O(3)$, and the pair $(\ell,\pi)$ labels its irreducible representations. We write $\mathcal V_\ell^\pi$ for the corresponding $(2\ell+1)$-dimensional space, whose elements $\mathbf t^{(\ell,\pi)}\in\mathcal V_\ell^\pi$ are spherical tensors with components $t_m^{(\ell,\pi)}$, $m=-\ell,\ldots,\ell$.

We call a tensor \emph{proper} when its parity matches the natural parity of its degree, and \emph{pseudo} when it carries the opposite parity:
\begin{equation}
\tilde\pi^{\mathrm{pr}}_\ell=(-1)^\ell,
\qquad
\tilde\pi^{\mathit{ps}}_\ell=-\tilde\pi^{\mathrm{pr}}_\ell=(-1)^{\ell+1}.
\end{equation}
Thus $\mathcal V_0^+$ contains proper scalars and $\mathcal V_1^-$ proper (polar) vectors, whereas $\mathcal V_1^+$ contains even pseudovectors (axial vectors) and $\mathcal V_0^-$ pseudoscalars.\footnote{A pseudoscalar is invariant under proper rotations but changes sign under inversion. A proper scalar is invariant under both.} The MLIP channels supplied to the chiral encoder block are proper tensors~\cite{Batzner2022, Batatia2022}. Our goal is to build a pseudoscalar, an element of $\mathcal V_0^-$, from these proper inputs. We first recall the tensor product that couples two such tensors, then show that a single product cannot produce pseudoscalars and that two successive products can.

\subsubsection{The tensor product}

Two spherical tensors $\mathbf t_1^{(\ell_1,\pi_1)}$ and $\mathbf t_2^{(\ell_2,\pi_2)}$ form the tensor product space,
\begin{equation}
    \mathcal V_{\ell_1}^{\pi_1}\otimes\mathcal V_{\ell_2}^{\pi_2},
\end{equation}
of dimension $(2\ell_1+1)(2\ell_2+1)$. This space decomposes as a direct sum of irreducible representations of $O(3)$,
\begin{equation}
    \mathcal V_{\ell_1}^{\pi_1}\otimes\mathcal V_{\ell_2}^{\pi_2} = \bigoplus_{\ell_3=|\ell_1-\ell_2|}^{\ell_1+\ell_2}
\mathcal V_{\ell_3}^{\pi_1\pi_2},
\label{eq:direct-sum}
\end{equation}
reached by an orthogonal change of basis from the products $\mathbf t_1^{(\ell_1,\pi_1)}\otimes\mathbf t_2^{(\ell_2,\pi_2)}$ to the degree-$\ell_3$ components, such that under rotation these components mix only within each degree-$\ell_3$ subspace and never between subspaces of different degree (Equation~\ref{eq:wigner-d}). The change of basis is given by the Clebsch--Gordan coefficients, so that the degree-$\ell_3$ output tensor $\mathbf t_3^{(\ell_3,\pi_3)}$ reads
\begin{equation}
t_{3,\,m_3}^{(\ell_3,\pi_3)}
=\sum_{m_1,m_2}
C^{\ell_3 m_3}_{\ell_1 m_1,\ell_2 m_2}\,
t_{1,\,m_1}^{(\ell_1,\pi_1)}\,
t_{2,\,m_2}^{(\ell_2,\pi_2)},
\label{eq:cg-contraction}
\end{equation}
where the Clebsch--Gordan coefficients $C^{\ell_3 m_3}_{\ell_1 m_1,\ell_2 m_2}$ are nonzero only when the selection rules are met:
\begin{align}
&\text{(degree)}\quad |\ell_1-\ell_2|\leq\ell_3\leq\ell_1+\ell_2, \label{eq:triangle}\\
&\text{(order)}\quad m_1+m_2=m_3. \label{eq:order}
\end{align}
The degree rule~\eqref{eq:triangle} is often referred to as the triangle inequality. We write $[\,\cdot\,]^{(\ell_3)}$ for the projection onto the degree-$\ell_3$ summand of Equation~\ref{eq:direct-sum}, i.e. the selection of the degree-$\ell_3$ component of a tensor product. To obtain pseudoscalars we apply two tensor products.

\subsubsection{Constructing pseudotensors}
Let $\mathbf a\in\mathcal V_{\ell_a}^{\pi_a}$ and $\mathbf b\in\mathcal V_{\ell_b}^{\pi_b}$ be two proper input tensors obtained from the latent features of an MLIP model. Applying the tensor product~\eqref{eq:cg-contraction} we obtain the intermediate,
\begin{equation}
    \textstyle\bigoplus_L \mathbf{x}^{(L,\pi_a\pi_b)} = \mathbf{a}^{(\ell_a,\pi_a)} \otimes \mathbf{b}^{(\ell_b,\pi_b)},
\end{equation}
where we define $\otimes$ as the spherical tensor product containing the change of basis, such that $L$ spans all admissible degrees $|\ell_a-\ell_b|\leq L\leq\ell_a+\ell_b$.

This product can already produce higher-order pseudotensors, whenever the output parity $\pi_a\pi_b$ is opposite to the natural parity $(-1)^L$ of its degree. Hence, when $\ell_a+\ell_b+L$ is odd, the intermediate, $\mathbf{x}$, is a pseudotensor. However, the triangle inequality prevents us from constructing a scalar ($L=0$) pseudotensor directly, which would require $\ell_a=\ell_b$ via the triangle selection rule. For proper inputs this forces the parity $(-1)^{\ell_a+\ell_b}=(-1)^{2\ell_a}=+$ to be even, and the result is a proper scalar. A single product therefore yields, at best, a pseudotensor of nonzero degree. A second product combines these higher-degree pseudotensors with a proper tensor to create a pseudoscalar.

\subsubsection{Two products make a pseudoscalar}
\label{sec:app-proof-pseudo}

We couple the intermediate pseudotensor to a third proper tensor $\mathbf c\in\mathcal V_{\ell_c}^{\pi_c}$ and project onto degree zero. By the triangle inequality~\eqref{eq:triangle}, degree zero is reached only when the two coupled degrees are equal,
\begin{equation}
L=\ell_c,
\end{equation}
so the intermediate degree is fixed to $\ell_c$. The resulting scalar
\begin{equation}
\chi=\left[\left[\mathbf a\otimes\mathbf b\right]^{(\ell_c)}\otimes\mathbf c \right]^{(0)}
\in\mathcal V_0^{\pi_a\pi_b\pi_c}
\end{equation}
is invariant under proper rotations, and its parity is $\pi_a\pi_b\pi_c$. By definition, it is a pseudoscalar precisely when
\begin{equation}
\pi_a\pi_b\pi_c=-1.
\end{equation}
For proper inputs, $\pi_i=(-1)^{\ell_i}$, this becomes $(-1)^{\ell_a+\ell_b+\ell_c}=-1$. Hence, a triple of proper tensors couples to a pseudoscalar exactly when the degrees satisfy the triangle inequality~\eqref{eq:triangle},
\begin{equation}
\begin{aligned}
    |\ell_a-\ell_b|\leq \ell_c\leq\ell_a+\ell_b, &\quad  \text{and}\\
    \ell_a+\ell_b+\ell_c,  &\quad \text{is odd}.
\end{aligned}
\label{eq:pseudoscalar-selection-rule}
\end{equation}

The simplest such example path is $\ell_a=\ell_b=\ell_c=1$, shown in the left panel of \figref{fig:method_figure}{b}. In \texttt{e3nn} parity notation it reads
\begin{equation}
1o\otimes 1o\rightarrow 1e,
\qquad
1e\otimes 1o\rightarrow 0o,
\end{equation}
where the first product builds a pseudovector from $\mathbf a$ and $\mathbf b$, and the second contracts it with the proper vector $\mathbf c$. Here $e$ and $o$ refer to even and odd tensors respectively. Up to sign and normalisation conventions, this is the scalar triple product
\begin{equation}
\mathbf c\cdot(\mathbf a\times\mathbf b).
\end{equation}

\begin{figure}[tb]
    \centering
    \includegraphics[width=0.9\linewidth]{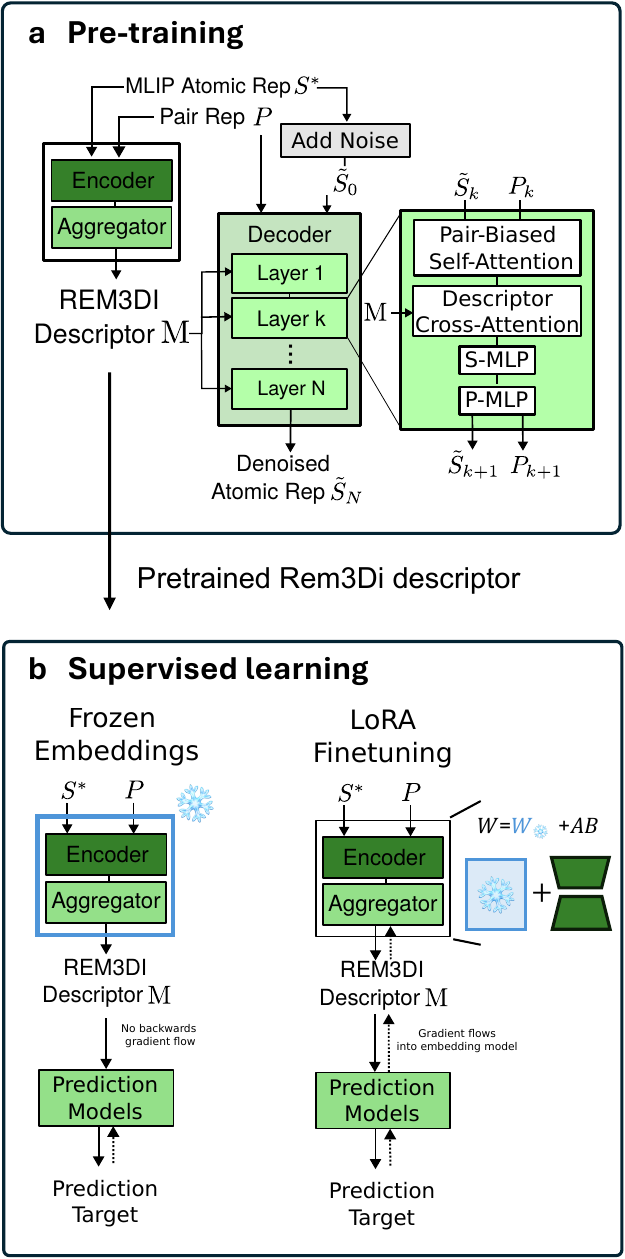}
    \caption{\textbf{Denoising pretraining and fine-tuning.} (a) The self-supervised training objective, in which the decoder reconstructs clean atomic representations from noisy ones using a clean global Rem3Di descriptor. (b) The supervised adaptation strategies, in which the pretrained encoder is either kept frozen or augmented with LoRA adapters while the task head is trained.}
    \label{fig:training}
\end{figure}

In practice $\mathbf a$, $\mathbf b$, and $\mathbf c$ are learnable linear mixings of the internal latent features $h_{\ell m}$, formed with parity- and rotation-preserving maps in \texttt{e3nn}~\cite{Geiger2022}. For features up to a maximum degree there are many combinations $(\ell_a,\ell_b,\ell_c)$ satisfying the selection rules, each such a choice is a \emph{path}. We take a learnable combination of paths, giving every path an independent multiplicative weight. An \texttt{e3nn} implementation of the construction is released with the code (Section~\ref{sec:data-avail}).

The resulting learnable pseudoscalars are invariant under proper rotations, change sign under inversion or reflection, vanish when the three tensors are coplanar, and take opposite signs for mirror-related configurations. To weight their contributions, they are gated by a factor computed from the frozen MLIP scalar features and projected by a bias-free linear map to form the chiral embeddings. These are appended to the invariant atomic features, as visualised in \figref{fig:method_figure}{b}.

\subsection{Pair representation and geometry}

MLIP features describe the many-body local environment of each atom. To facilitate learning longer-range dependencies, Rem3Di supplements the atom representation $S\in\mathbb{R}^{N\times d_\text{model}}$ with a pair representation $P\in\mathbb{R}^{N\times N\times d_\text{pair}}$ that explicitly encodes the geometry and evolving chemical relationship between every pair of atoms. As shown in \figref{fig:method_figure}{c}, $P$ is initialised from the interatomic distances
\begin{equation}
d_{ij}=\lVert \vec r_i-\vec r_j\rVert,
\end{equation}
which are expanded using a DimeNet-style Bessel basis,
\begin{equation}
e_n(d)
=
\sqrt{\frac{2}{c}}\,
\frac{\sin(n\pi d/c)}{d}.
\end{equation}
The resulting radial features are passed through a bias-free linear projection, symmetrised over the two atom indices, and masked such that pairs involving padded atoms are zero. We use a large cutoff of $c=32$~\AA{} with 16 basis functions, allowing the geometric encoding to span large parts of a molecule rather than only bonded or spatially adjacent atoms. Optionally, additional pairwise information may be supplied by a two-dimensional molecular graph or a $k$-nearest-neighbour graph constructed in three-dimensional space.

The global encoder, illustrated in \figref{fig:method_figure}{d}, jointly updates the atom and pair representations using a stack of transformer blocks. Each block comprises pair-biased multi-head self-attention, an atom feed-forward update, a pair update, and a pair feed-forward update. Because molecules are unordered sets of atoms, no sequence positional embeddings are used. Permutation equivariance is instead maintained by applying identical operations to all atoms and atom pairs, together with padding masks for variable-sized molecules.

As in AlphaFold2\cite{jumper}, at each layer, the current pair representation enters the atom update as an additive, head-specific bias to the self-attention logits,
\begin{equation}
\Delta S
=
\mathrm{softmax}\left(
\frac{QK^{\top}}{\sqrt{d_k}} + B
\right)V,
\end{equation}
where the bias $B$ depends on the pair representation at the current layer, such that for each attention head $h$ and atom pair $(i,j)$,
\begin{equation}
B_{ij}^{h} 
=
w_h^{\top}\,
\mathrm{LayerNorm}(P_{ij}),
\end{equation}
where $w_h$ is a bias-free linear map producing one scalar bias for attention head $h$ and atom pair $(i,j)$. Consequently, the attention assigned between two atoms can depend on their current geometric and chemical relationship, rather than only on the similarity of their atom embeddings.

The pair representation is itself refined at every encoder layer using the current pair features, the embeddings of the two corresponding atoms, and the original radial distance encoding, as visualised in \figref{fig:method_figure}{d}. This bidirectional coupling allows the atom representations to be learned in the context of the full molecular geometry while the pair pathway progressively develops representations of relationships such as covalent bonding, conjugation, electrostatic interactions, and non-covalent contacts. The original radial features are retained throughout the encoder, providing a fixed geometric reference for the learned pair updates.

\begin{figure*}[tb]
    \centering
    \includegraphics[width=0.97799\linewidth]{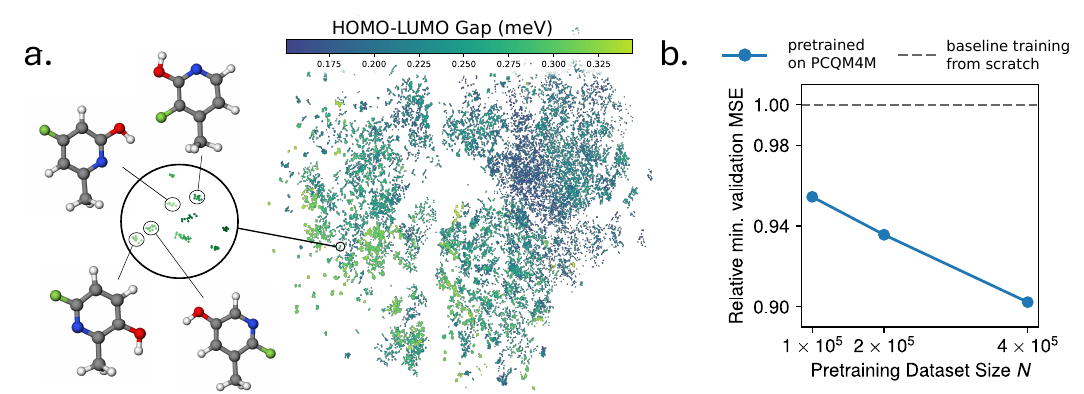}
    \caption{\textbf{Denoising pretraining.} (a) UMAP of the resulting pretrained molecular representations, with insets showing that chemically similar molecules are clustered together. (b) Performance on a HOMO-LUMO gap regression task, on a 10,000-molecule subset of QM9, as a function of pretraining dataset size.}
    \label{fig:denoise}
\end{figure*}

Local positional and chemical information is already encoded in the MLIP features through their short-range many-body environments. Combining these features with all-pairs geometric attention allows Rem3Di to model longer-range effects, including conjugation, electrostatics, and \mbox{$\pi$--$\pi$} stacking, and to learn distance-dependent weighting of atomic environments before aggregation. Constructing and updating the pair representation requires $\mathcal{O}(N^2)$ operations and memory, but this does not change the asymptotic complexity of the encoder, which is already $\mathcal{O}(N^2)$ owing to full self-attention. For substantially larger systems, the same architecture could be adapted by imposing a global cutoff or using sparse attention, or by replacing the all-pairs encoding with a formulation that recovers global atomic representations at linear cost~\cite{Frank2026}.

After the final encoder layer, the variable-sized set of atom representations is converted into a fixed-dimensional molecular descriptor using permutation-invariant pooling by multi-head attention (PMA)~\cite{Lee2018}. A fixed number of learned seed vectors cross-attend to the final atom representations, producing a fixed number of pooled tokens independent of molecular size. These tokens are subsequently flattened or projected to obtain the final Rem3Di descriptor. We ablate the number of learned query vectors and use PMA unless otherwise stated.

\subsection{Training of Rem3Di}

We first pretrain Rem3Di in a self-supervised way, requiring no experimentally labelled data. Subsequently, we fine-tune the model for specific tasks.

\subsubsection{Denoising pretraining}
\label{sec:pretraining}

Experimental labels for pharmaceutically relevant molecules are scarce and often proprietary, whereas unlabelled molecules are available at scale through databases such as ZINC, PubChem and ChEMBL~\cite{Irwin2012, Kim2024, Mendez2018}. Additionally, there are large databases of plausible SMILES strings, for which 3D conformers can be generated with RDKit and relaxed in batch using an MLIP~\cite{Cohen2025, Batatia2024}. To exploit these collections, we pretrain Rem3Di with a \emph{self-supervised} objective that requires no labels. The principle, shared with denoising autoencoders and diffusion models~\cite{Ho2020a, Song2020}, is to corrupt the input and train the model to undo the corruption. Succeeding at this forces the model to learn a useful representation. In atomistic machine learning, denoising has been used to pretrain force fields~\cite{Liao2024, Liu2022} and property predictors~\cite{Zaidi2022, Wijaya2024}, often by perturbing atomic {coordinates}, which can create high-energy, bond-breaking geometries~\cite{Feng2023, Liu2025}. Furthermore, generative models have been trained to denoise in the latent space of variational autoencoders for molecular and material structures~\cite{Joshi2025, Park2026}, and to run joint flow matching over real-space coordinates and per-residue latent variables for atomistic protein generation~\cite{Geffner2025, Didi2026}.

We add noise direclty in pretrained MLIP {feature space}. As illustrated in \figref{fig:training}{a}, the encoder first maps the clean molecule to its descriptor $M$. The preprocessed per-atom features are then corrupted with isotropic Gaussian noise of fixed scale $\sigma$, and a transformer decoder is trained to reconstruct the {clean} features from the noised ones. Crucially, the only trainable global information path from the clean molecule to
the decoder is through cross-attention to the descriptor $M$, while the pair representation $\mathbf{P}$ is provided only as detached geometric context. For the reconstruction to succeed, $M$ must therefore carry a faithful, compressed summary of the whole molecule: it is an information bottleneck. The objective is a padding-masked mean-squared error, rescaled by the corruption variance so its scale is comparable across noise levels,
\begin{equation}
\mathcal{L}_\text{denoise} = \frac{1}{N_\text{atoms}\,\sigma^{2}} \sum_{i\in\text{real}} \big\lVert \hat{S}^{0}_i - S^{0}_i \big\rVert^{2},
\end{equation}
where $S^{0}$ are the clean preprocessed atom features and $\hat{S}^{0}$ the decoder reconstructions. To discourage representational collapse, in which every molecule maps to the same vector, we optionally add VICReg variance and covariance regularisers on the descriptor batch~\cite{Bardes2021}: a hinge that keeps each dimension's standard deviation above a target, and an off-diagonal covariance penalty that decorrelates dimensions, with the denoising term itself playing the role of VICReg's invariance term. We optimise all trainable parameters (preprocessor, encoder and decoder) with AdamW under a one-cycle learning-rate schedule, while the foundation MLIP stays frozen. After pretraining the decoder is discarded and the encoder produces the descriptors. Full hyperparameters and pseudocode are given in Appendices~\ref{app:hyperparameters} and~\ref{app:pretrain-pseudocode}.

\begin{figure*}[tb]
    \centering
    \includegraphics[width=1 \linewidth]{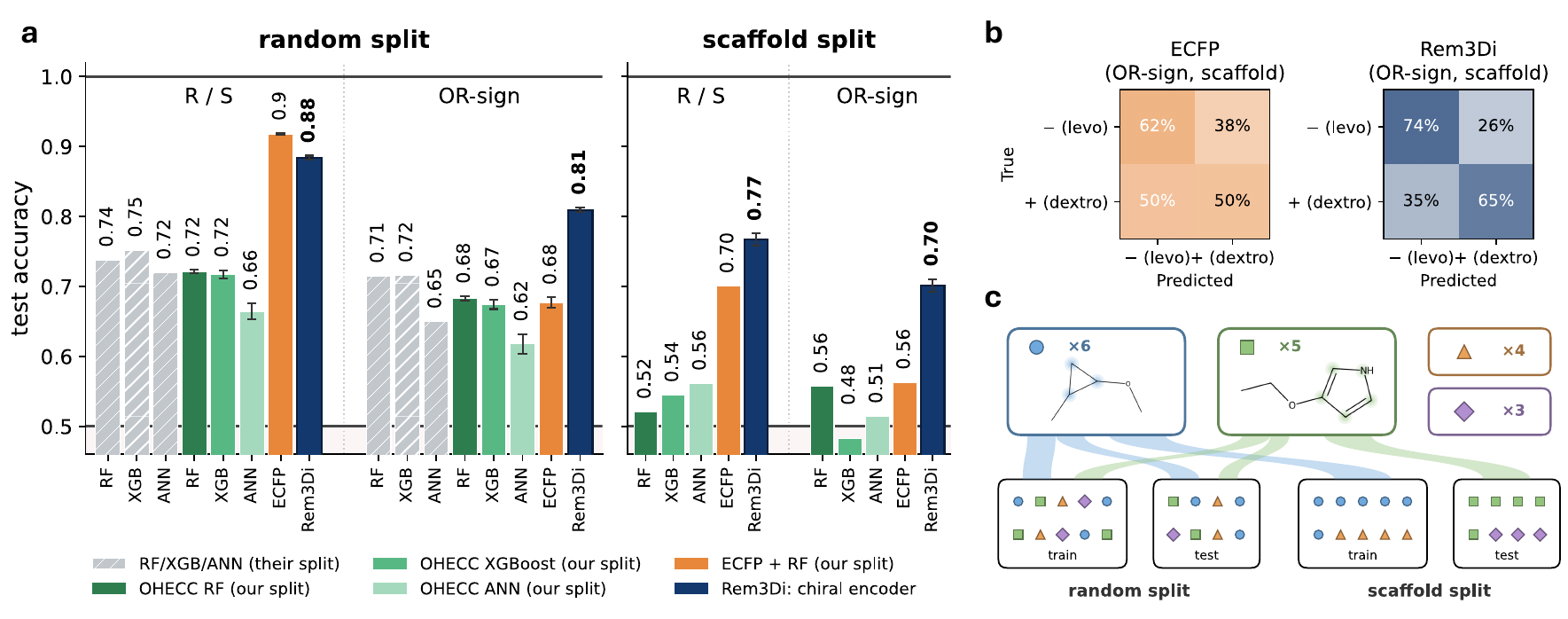}
    \caption{\textbf{Chirality property prediction under random and scaffold splits on QM9-OR.} (a) Test accuracy of six methods on two targets, R/S (handedness of chiral centres) and OR-sign (sign of the optical rotation at $589.3$~nm), under a random split and a Bemis--Murcko scaffold split. The test set is mirror-balanced, so chance accuracy is $0.50$ (dashed line). Grey hatched bars are the OHECC descriptor baselines of Zhou et al.~\cite{Zhou2025} on their own random split. Coloured bars are the methods retrained on our splits: random forest, XGBoost and ANN on the OHECC descriptor, a Morgan-fingerprint random forest (ECFP), and the chiral encoder Rem3Di. Error bars are $\pm1$~s.d.\ over four seeds. (b) Row-normalised confusion matrices for OR-sign on the scaffold split, on single-stereocentre test molecules ($n=1229$), for Rem3Di (balanced accuracy $\approx0.70$) and ECFP ($\approx0.56$). (c) Schematic of the train/test split: molecules sharing a scaffold are highly correlated, so a random split places members of one cluster in both train and test, which leaks information from training into the test set~\cite{Wu2018, Sheridan2013}, while a scaffold split reduces data leakage.}
    \label{fig:results_chiral}
\end{figure*}

\subsubsection{Supervised training for property prediction}

After pretraining (\figref{fig:training}{a}), Rem3Di is adapted to labelled property-prediction tasks. Throughout, the foundation MLIP is kept frozen and only the components downstream of it are trained or adapted. We consider three training strategies of increasing capacity. First, in the \emph{frozen-descriptor} strategy the encoder and aggregator are fixed and the descriptor is used as an input feature for a separate predictor, either a lightweight task head or a classical regressor fit under repeated nested cross-validation with randomised hyperparameter search~\cite{Ash2024}. The classical regressors we use are ridge regression~\cite{Hilt1977}, random forests~\cite{Breiman2001} and gradient-boosted trees. This isolates the quality of the learned descriptor from any task-specific adaptation of the representation.

\begin{table*}[tb]
\centering
\caption{\textbf{Drug-property benchmark on TDC and MoleculeNet.} Per-task results shown as two stacked blocks: classification (top, AUROC $\uparrow$, higher is better) and regression (bottom, lower is better $\downarrow$, TDC scored by MAE, MoleculeNet by RMSE). Cells: mean\dev{std} over seeds (our rows: 3 seeds, external baselines as published). Within each block, external published baselines (top) and matched-protocol evaluations (bottom) are separated by a rule. In each column the \colorbox{firsthl}{best} and \colorbox{secondhl}{second-best} rankable values are shaded (ties share a rank). ``--'': endpoint not reported for that model. $^{\dagger}$: coverage $<90\%$ (shown but excluded from ranking). Fine-tuned Rem3Di is the best matched-protocol model on all six regression endpoints. The complete leaderboard, including the MoleculeNet classification suite, is given in Appendix~\ref{app:benchmark} (Table~\ref{tab:benchmark-full}).}
\vspace{2 mm}
\label{tab:benchmark}
\scriptsize
\renewcommand{\arraystretch}{1.15}
\setlength{\tabcolsep}{5pt}
\begin{tabular}{l cccccc}
\toprule
\textsc{Classification, AUROC} $\uparrow$ & \textsc{BBB} & \textsc{HIA} & \textsc{Pgp} & \textsc{Bioav.} & \textsc{Tox-Avg} & \textsc{CYP-Avg} \\
\midrule
D-MPNN / Chemprop & 0.864\dev{.010} & \cellcolor{firsthl}0.976\dev{.004} & 0.889\dev{.005} & 0.617\dev{.050} & 0.821\dev{.019} & 0.819\dev{.004} \\
AttentiveFP & 0.855\dev{.011} & \cellcolor{secondhl}0.974\dev{.007} & 0.892\dev{.012} & \cellcolor{secondhl}0.632\dev{.039} & 0.842\dev{.010} & 0.749\dev{.008} \\
DeepMol & 0.774\dev{.023} & 0.880\dev{.012} & 0.821\dev{.007} & 0.509\dev{.026} & 0.735\dev{.015} & 0.770\dev{.008} \\
FPGNN & 0.888\dev{.018} & 0.958\dev{.012} & \cellcolor{firsthl}0.930\dev{.007} & \cellcolor{firsthl}0.666\dev{.035} & 0.860\dev{.017} & \cellcolor{secondhl}0.866\dev{.004} \\
TranFoxMol & 0.868\dev{.019} & 0.951\dev{.036} & 0.875\dev{.011} & 0.619\dev{.019} & 0.837\dev{.017} & 0.860\dev{.006} \\
\midrule
ECFP4 + LightGBM & 0.884\dev{.005} & 0.905\dev{.048} & 0.907\dev{.013} & 0.591\dev{.036} & 0.829\dev{.013} & \cellcolor{secondhl}0.866 \\
UniMol2 (frozen) & 0.884\dev{.014} & 0.902\dev{.088}$^{\dagger}$ & 0.898\dev{.017} & 0.617\dev{.041} & 0.833\dev{.020}$^{\dagger}$ & \cellcolor{firsthl}0.867 \\
MuMo (frozen) & 0.893\dev{.019} & 0.980\dev{.012}$^{\dagger}$ & 0.891\dev{.006} & 0.626\dev{.048} & 0.812\dev{.012}$^{\dagger}$ & 0.852 \\
\textbf{Rem3Di (frozen)} & \cellcolor{secondhl}0.894\dev{.004} & 0.949\dev{.019}$^{\dagger}$ & 0.898\dev{.012} & 0.558\dev{.057} & \cellcolor{secondhl}0.863\dev{.013} & 0.862 \\
\textbf{Rem3Di (fine-tuned)} & \cellcolor{firsthl}0.898\dev{.007} & 0.974\dev{.008}$^{\dagger}$ & \cellcolor{secondhl}0.911\dev{.006} & 0.561\dev{.018} & \cellcolor{firsthl}0.871\dev{.016} & \cellcolor{firsthl}0.867 \\
\midrule[\heavyrulewidth]
\textsc{Regression, error} $\downarrow$ & \multicolumn{4}{c}{\textsc{TDC, MAE}} & \multicolumn{2}{c}{\textsc{MoleculeNet, RMSE}} \\
\cmidrule(lr){2-5}\cmidrule(lr){6-7}
\textsc{Model} & \textsc{LD50} & \textsc{Caco-2} & \textsc{PPBR} & \textsc{LIPO} & \textsc{ESOL} & \textsc{FreeSolv} \\
\midrule
D-MPNN / Chemprop & \cellcolor{secondhl}0.607\dev{.022} & 0.388\dev{.077} & \cellcolor{secondhl}8.158\dev{.314} & \cellcolor{firsthl}0.448\dev{.014} & 1.050\dev{.008} & 2.082\dev{.082} \\
AttentiveFP & 0.678\dev{.012} & 0.401\dev{.032} & 9.373\dev{.335} & 0.572\dev{.007} & 0.877\dev{.029} & 2.073\dev{.183} \\
DeepMol & \cellcolor{firsthl}0.589\dev{.006} & 0.327\dev{.012} & 9.533\dev{.162} & 0.660\dev{.004} & -- & -- \\
GROVER & -- & -- & -- & -- & 0.831\dev{.120} & 1.544\dev{.397} \\
MolCLR & -- & -- & -- & -- & 1.271\dev{.040} & 2.594\dev{.249} \\
GraphMVP & -- & -- & -- & -- & 1.029\dev{.033} & -- \\
GEM & -- & -- & -- & -- & 0.798\dev{.029} & 1.877\dev{.094} \\
Uni-Mol & -- & -- & -- & -- & 0.788\dev{.029} & 1.480\dev{.048} \\
FPGNN & 0.638\dev{.024} & \cellcolor{secondhl}0.326\dev{.040} & 8.465\dev{1.709} & 0.544\dev{.011} & 0.658\dev{.006} & 1.106\dev{.195} \\
TranFoxMol & 0.645\dev{.036} & 0.487\dev{.068} & 9.055\dev{.523} & 0.525\dev{.024} & 0.930\dev{.261} & 1.225\dev{.155} \\
\midrule
ECFP4 + LightGBM & 0.659\dev{.005} & 0.477\dev{.011} & 9.646\dev{.122} & 0.627\dev{.014} & 1.069\dev{.124} & 2.178\dev{.400} \\
UniMol2 (frozen) & 0.682\dev{.012} & 0.338\dev{.018} & 8.985\dev{.244} & 0.602\dev{.017} & 0.655\dev{.065} & 1.169\dev{.144} \\
MuMo (frozen) & 0.693\dev{.030} & 0.353\dev{.022} & 8.265\dev{.287} & 0.677\dev{.008} & \cellcolor{secondhl}0.647\dev{.039} & 1.243\dev{.111} \\
\textbf{Rem3Di (frozen)} & 0.656\dev{.015} & 0.368\dev{.022} & 8.536\dev{.313} & 0.625\dev{.006} & 0.655\dev{.071} & \cellcolor{secondhl}1.069\dev{.205} \\
\textbf{Rem3Di (fine-tuned)} & 0.638\dev{.031} & \cellcolor{firsthl}0.326\dev{.016} & \cellcolor{firsthl}8.054\dev{.237} & \cellcolor{secondhl}0.499\dev{.015} & \cellcolor{firsthl}0.594\dev{.057} & \cellcolor{firsthl}0.820\dev{.177} \\
\bottomrule
\end{tabular}
\end{table*}

Second, we use parameter-efficient LoRA fine-tuning~\cite{Hu2021}. In this setting, the pretrained Rem3Di encoder weights remain frozen but are augmented with trainable low-rank updates, while the supervised prediction head is trained in the usual way. For a linear projection with pretrained weight $W \in \mathbb{R}^{d_{\mathrm{out}}\times d_{\mathrm{in}}}$, LoRA defines the effective weight as
\begin{equation}
W_{\mathrm{LoRA}} = W + \Delta W, \qquad \Delta W = \frac{\alpha}{r}\,BA,
\end{equation}
with $A \in \mathbb{R}^{r\times d_{\mathrm{in}}}$, $B \in \mathbb{R}^{d_{\mathrm{out}}\times r}$ and rank $r \ll \min(d_{\mathrm{out}}, d_{\mathrm{in}})$, so that only $r(d_\text{in}+d_\text{out})$ parameters are trained per adapted layer (in our experiments we insert rank-$16$ adapters, scaled by $\alpha=32$, into the encoder attention and pooling projections). Restricting adaptation to a low-rank subspace is well suited to the small-label regime typical of molecular property prediction, where unconstrained fine-tuning readily overfits. Third, as a high-capacity baseline, we \emph{fully fine-tune} all trainable Rem3Di weights downstream of the frozen MLIP together with the prediction head.

\section{Results}

We investigate the Rem3Di molecular descriptor from its semantic nature to its performance on cheminformatics regression and classification tasks. We find that Rem3Di maintains the transferable nature of the underlying MLIP and, under a matched evaluation protocol, is the best model on the regression endpoints of widely used drug-property benchmarks, with its advantage most visible on tasks governed by 3D physics. We then apply Rem3Di in a setting where the covalent graph that conventional bond-based descriptors assume is not well defined, namely transition-metal complexes.

\begin{figure*}[tb]
    \centering
    \includegraphics[width=0.847799\linewidth]{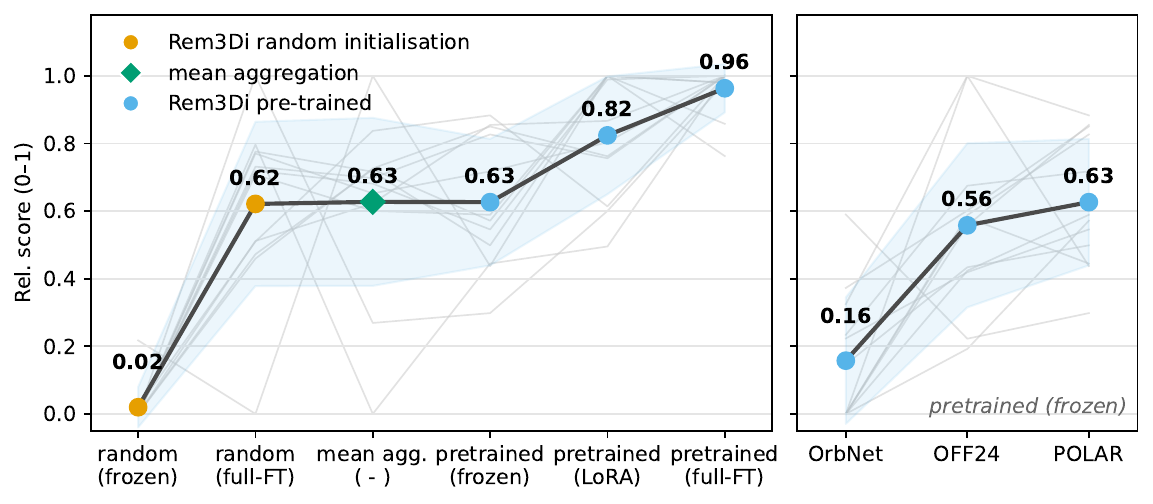}
    \caption{\textbf{Rem3Di architectural and training ablations.} Relative performance across 12 drug-property tasks. Left: encoder initialisation (random or denoising-pretrained) crossed with the adaptation strategy (frozen descriptor or full fine-tune), together with a parameter-free mean-aggregation baseline. Right: the same comparison across three frozen foundation models. For each task the native metric is oriented so that higher is always better (error metrics are sign-flipped) and then min--max normalised to a relative score, where 0 is the worst and 1 the best of the conditions for that task. Grey lines show the 12 individual tasks, and the mean over all 12 as a thicker blue line, with the shaded band denoting $\pm1$ standard deviation across tasks and annotated values giving the mean at each condition. Pretraining followed by full fine-tuning gives the highest score ($0.96$), the pretrained frozen descriptor ($0.63$) matches both the randomly initialised, fully fine-tuned model ($0.62$) and the mean-aggregation baseline ($0.63$).}
    \label{fig:finetuning_ablation}
\end{figure*}

\subsection{Denoising pretraining structures the molecular representation}

The self-supervised pretraining protocol produces chemically structured descriptors and improves performance on regression tasks.
We show that a Rem3Di model pretrained on the PCQM4M~\cite{Nakata2023} dataset, using the denoising objective of \figref{fig:training}{a}, yields chemically structured descriptors for molecules in the QM9 dataset. \figref{fig:denoise}{a} shows clusters in the low-dimensional UMAP projection. The molecules within an example cluster contain the same functional groups, but in differently ordered substitutions along the aromatic ring. Conformers of each molecule, obtained by short ($200$~fs) molecular dynamics simulation, fall in the same region of descriptor space, as indicated by the multiple encircled scatter points, showing that the descriptor is stable against small geometric perturbations. This hierarchical structure of the Rem3Di descriptor improves the performance on regression tasks compared to randomly initialised weights, as outlined in Section~\ref{sec:results-ablation-pre-training}. Consistent with this, downstream accuracy on a HOMO-LUMO gap regression task grows with the amount of unlabelled data used for pretraining (\figref{fig:denoise}{b}).

\subsection{Chiral encoder for optical rotation predictions}

A chiral molecule rotates the plane of polarised light, and its two enantiomers rotate it by equal amounts in opposite directions. The sign of this optical rotation (OR) therefore records which enantiomer a molecule is. This makes OR prediction a suitable and controlled benchmark for chirality prediction. We compare Rem3Di's chiral encoder, which builds the pseudoscalar features illustrated in \figref{fig:method_figure}{b}, with alternative descriptors on QM9-OR~\cite{Zhou2025}, a dataset that pairs $121{,}416$ molecules from QM9~\cite{Ramakrishnan2014} ($\leq 9$ heavy atoms) with optical rotations computed by TD-DFT (CAM-B3LYP/6-31G**) and absolute R/S (CIP) labels. We train two binary heads on the frozen descriptor, one for the configuration of the stereocentre (R/S, assigned by the Cahn--Ingold--Prelog rules) and one for the sign of the optical rotation at $589.3$~nm (OR-sign). Each molecule is included with its mirror image and the labels flipped, so the two classes are balanced and chance accuracy is $0.50$.

We compare Rem3Di with the OHECC descriptor classifiers of Zhou et al.\ (random forest, XGBoost, ANN)~\cite{Zhou2025} and a chirality-aware Morgan fingerprint with a random forest (ECFP)~\cite{Rogers2010}, under a random split and a Bemis--Murcko scaffold split~\cite{Bemis1996}. The scaffold split groups molecules by their Bemis-Murcko framework and assigns each group entirely to either the training or the test set.
As visualised in \figref{fig:results_chiral}{c}, a random split places molecules that share a scaffold in both train and test, so a method can score highly by matching a test molecule to a near-duplicate in the training set, a form of data leakage~\cite{Wu2018, Sheridan2013}. A scaffold split does not place the same scaffold in the train and test set, thereby reducing data leakage.

We find that across the benchmarks all models perform better at predicting the handedness of stereocentres (R/S) than the sign of the optical rotation (OR-sign). Across both the random and the scaffold split, Rem3Di performs better than the OHECC and ECFP at predicting OR-sign.
On the random split, ECFP reaches $0.92$ on R/S, above Rem3Di at $0.88$, because the Morgan chirality bit records the CIP label and R/S is hence easy to regress. This does not transfer to the scaffold split, where ECFP drops to $0.70$ on R/S and the OHECC baselines to $0.52$--$0.56$. For OR-sign, which has no 2D correlate, ECFP scores $0.68$ on the random split and $0.56$ on the scaffold split, and the OHECC baselines are close to chance. Rem3Di stays above chance on the scaffold split for both targets (R/S $0.77$, OR-sign $0.70$) and has the highest OR-sign accuracy on both splits (\figref{fig:results_chiral}{a}).

The confusion matrices for OR-sign on the scaffold split (\figref{fig:results_chiral}{b}) separate accuracy from class bias. Rem3Di predicts both signs at similar rates (balanced accuracy $\approx 0.70$). ECFP predicts one sign preferentially, with balanced accuracy $\approx 0.56$, consistent with its OR-sign bar in panel (a).

\subsection{Property prediction in drug discovery}

We benchmark Rem3Di against published baselines on two public drug-property suites, the Therapeutics Data Commons (TDC) ADMET collection~\cite{Huang2021} and MoleculeNet~\cite{Wu2018}, both as a frozen descriptor and after task-specific fine-tuning. Table~\ref{tab:benchmark} reports the TDC ADMET classification and the combined TDC/MoleculeNet regression results. The complete per-task leaderboard with all baselines, including the MoleculeNet classification suite, is given in Appendix~\ref{app:benchmark}. 
These benchmarks include experimentally grounded pharmacokinetic and safety endpoints, such as blood--brain-barrier penetration, intestinal absorption, P-glycoprotein transport, oral bioavailability, cytochrome P450 metabolism, toxisity plasma protein binding, Caco-2 permeability, lipophilicity, and acute toxicity, together with physically well-defined solubility and hydration-free-energy tasks from MoleculeNet. The resulting tasks are challenging because they combine small labelled datasets, assay noise, endpoint heterogeneity, class imbalance, and scaffold-level generalisation, but they remain highly informative because performance reflects molecular features directly relevant to drug discovery. The same difficulty is reflected in community blind challenges on pharmaceutical data~\cite{MacDermottOpeskin2025}, where held-out experimental endpoints remain hard to predict.

To obtain a controlled comparison, we evaluate a set of baselines under one matched protocol, with the same train/test splits and regression heads. As a classical cheminformatics baseline we use an ECFP4\,+\,LightGBM 2D anchor~\cite{Rogers2010}, and the frozen UniMol2 and MuMo descriptors as machine-learned baselines. Across the twelve endpoints shown here, fine-tuned Rem3Di is the best model in the matched-protocol band on ten (all six regression endpoints and four of the six TDC classification endpoints), and the single best entry overall, including the external published baselines, on seven. The advantage is concentrated in regression: fine-tuned Rem3Di is the best matched-protocol model on \emph{all six} regression endpoints (ESOL, FreeSolv, Caco-2, LD50, Lipophilicity, PPBR) and the best entry overall on four of them. The separation is largest on tasks governed by 3D physics, most clearly the solvation free energy FreeSolv, where Rem3Di improves on every 2D and graph baseline and on its own frozen descriptor (RMSE $1.07\!\rightarrow\!0.82$). On classification it leads the TDC blood--brain-barrier, toxicity and CYP endpoints.

For most of the tasks where fine-tuned Rem3Di ranks first, there are similarly performing models within one standard deviation of cross-seed variation.
The robust finding is one of breadth. Rem3Di is simultaneously at or near the best performing models across the entire regression family and most TDC classification endpoints, a pattern that no single baseline reproduces. Rem3Di is not uniformly strong. It is weakest on TDC bioavailability, where several baselines do better. More generally, on the smallest classification datasets full fine-tuning can \emph{degrade} performance relative to the frozen descriptor while improving every regression endpoint, a pattern we analyse in more detail on the MoleculeNet classification suite in Appendix~\ref{app:benchmark}. This contrast motivates the ablation studies below.

\begin{figure*}
    \centering
    \includegraphics[width=.877\linewidth]{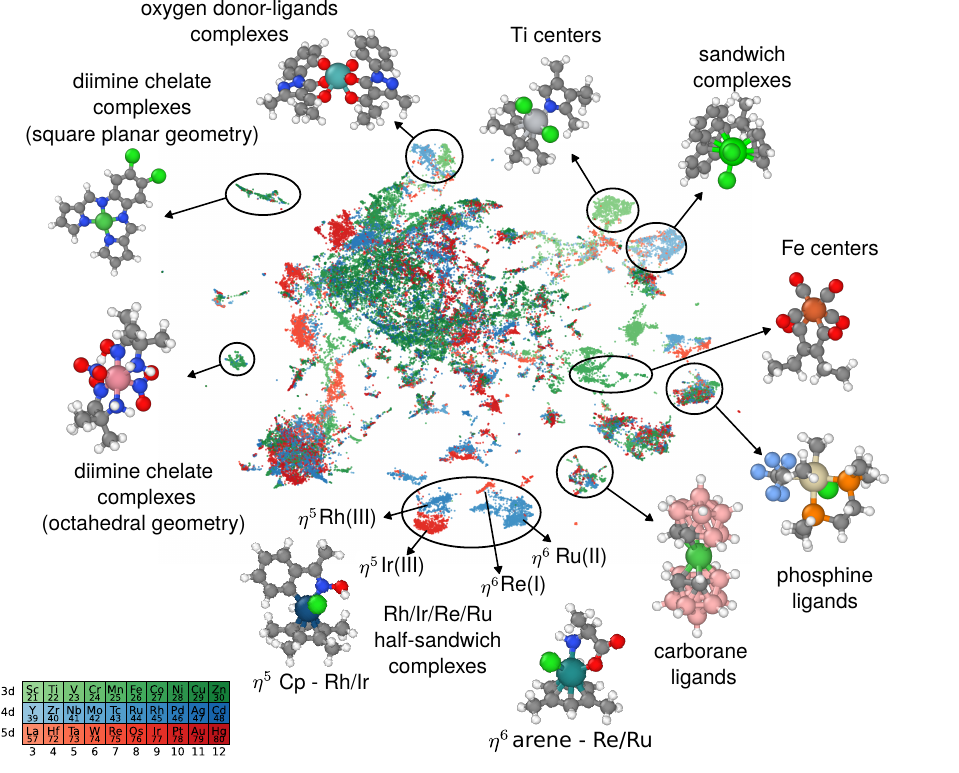}
    \caption{\textbf{Transition-metal complexes.} UMAP of the molecular descriptors after denoising pretraining without ground-truth labels. Insets highlight the clustering of chemically similar compounds, which groups TMCs by geometry, ligand chemistry, and metal centre identity.}
    \label{fig:tm-complexes}
\end{figure*}

\subsection{Ablations}

Having established Rem3Di's performance across a variety of benchmarks, we investigate key components of the architecture. We find that pretraining and fine-tuning are both required for competitive performance on benchmarks, and that Rem3Di's performance is strongly linked to the choice of frozen MLIP.

To capture ablation trends from noisy experimental datasets, we compare models across 12 drug-property tasks on their official splits. For each task we min--max normalise the results, such that higher is better and all scores are between 1 (best) and 0 (worst), as compared to all tested models. \figref{fig:finetuning_ablation}{} summarises the ablations. Their absolute values are given in Table~\ref{tab:benchmark-full}.

To keep the comparison controlled, every variant is pretrained with the
\emph{identical} denoising objective, optimiser schedule, and Rem3Di architecture
used for the main model (Section~\ref{sec:pretraining}). All models are trained on the same set of GEOM-drugs top-one conformers ($\sim$281k molecules), and the 3D conformer used for a given SMILES string is identical across variants.

\subsubsection{Pretraining and fine-tuning}
\label{sec:results-ablation-pre-training}

We find that the highest performance is obtained when pretraining the Rem3Di aggregator and subsequently fine-tuning the model. To quantify each contribution we perform the ablations visualised in \figref{fig:finetuning_ablation}{}. The figure shows the relative performance, min-max normalised, of Rem3Di models that are randomly initialised or pretrained. For each initialisation we fit models in which the molecular descriptor is frozen during fine-tuning and only the task-specific heads are trained, as well as models in which the entire architecture is fine-tuned. The lowest score, $0.02$, occurs for randomly initialised weights that are then kept frozen. Allowing those weights to vary during fine-tuning raises the score to $0.62$, against $0.96$ for the best model, which combines pretraining with full fine-tuning. Pretraining Rem3Di and then freezing the weights gives $0.63$, comparable to the randomly initialised model that is fully fine-tuned.

As a baseline, we mean aggregate the individual MLIP features, removing all learnable parameters from the Rem3Di descriptor. This parameter-free model scores $0.63$, matching the pretrained frozen descriptor. This ablation supports the initial hypothesis that MLIPs learn chemically rich descriptors, which can be used for tasks outside of direct simulation. The remaining gap to the pretrained and fully fine-tuned model, $0.63$ to $0.96$, is what the learned aggregation contributes.

\subsubsection{Base foundation model}

To identify how important the base-MLIP model is, we pretrain a set of Rem3Di models on two MACE models and one Orb model. \figref{fig:finetuning_ablation}{} (right) shows that MACE-POLAR-1-M scores highest ($0.63$), followed by MACE-OFF24 ($0.56$), with the Orb model lowest ($0.16$). MACE-POLAR-1-M is a polarisable electrostatic model~\cite{Batatia2026} trained on the OMol dataset~\cite{Levine2025}, whereas the MACE-OFF24 features were trained on SPICEv2~\cite{Eastman2024}. The results indicate that the base foundation model can have a large impact on the final performance.

\subsection{Descriptors for transition-metal complexes}

Finally, we evaluate Rem3Di on transition-metal complexes (TMCs), a chemically distinct setting in which conventional graph-based descriptors can be brittle because coordination and organometallic bonding often involve delocalised, polycentric, metal-ligand, and metal-metal interactions that are not naturally captured by standard molecular graphs~\cite{Brammer2022,Rasmussen2025}.
Hand-built cheminformatics descriptors typically assume a fixed covalent bonding pattern, which breaks down for TMCs: the metal-ligand interaction is usually coordinate (dative) bonding and can involve multi-centre $\pi$ interactions, so a unique bond order is ambiguous~\cite{Brammer2022}. Descriptors that depend on well-defined covalent graphs therefore struggle or require restrictive assumptions, such as a small set of metals and ligands and a few interatomic distances chosen a priori~\cite{Nandy2021, Heller2015,Rasmussen2025}. One common alternative is the revised autocorrelation functions (RACs), which encode the correlation of selected atomic properties around the metal centre~\cite{Janet2017}. This descriptor gap hinders the exploitation of transition-metal compounds as pharmaceutical agents, particularly as anti-cancer drugs~\cite{Meggers2009,Mjos2014}.

We probe whether a 3D, physics-informed representation can retain information that is largely absent from circular substructure fingerprints such as ECFP.
Rem3Di captures chemical trends in transition-metal complexes in the tmQM dataset~\cite{Balcells2020}. \figref{fig:tm-complexes}{} shows the UMAP projection of the Rem3Di descriptors after pretraining a model via the denoising objective of \figref{fig:training}{a} on the tmQM dataset. Clusters form based on the transition-metal centre species, the ligand chemistry, or the geometry. Specific clusters, such as the Rh/Ir/Re/Ru half-sandwich complexes, show that the model can also capture electronic interactions between different structural features of the complexes.
This shows that Rem3Di is transferable to a diverse set of systems that can be modelled with the underlying MLIP, without having to hand-pick relevant features for systems of interest.

Metal-containing compounds are largely absent from the screening libraries and pretraining corpora that drive small-molecule discovery~\cite{Adamczyk2025}, and metallodrug candidates have lacked a general-purpose descriptor to organise them~\cite{MedinaFranco2022,Anthony2020}. By resolving transition-metal complexes into a fixed-size representation ordered by metal centre, ligand chemistry, and geometry without predefined bonding rules, Rem3Di supplies the input that standard cheminformatics workflows require. Similarity search, descriptor-based QSAR regression, and virtual library screening, which are routine for organic small molecules but obstructed for coordination compounds by the ambiguity of the covalent graph, therefore become applicable to metallodrug space through the same descriptor evaluated here.

\section{Conclusion}
We introduced Rem3Di, a framework that repurposes the latent features of atomistic foundation models as fixed-size, 3D-aware molecular descriptors. Per-atom features from a frozen machine-learning interatomic potential are aggregated by a geometry-biased, permutation-invariant transformer into a single molecular embedding, pseudoscalar channels constructed from the equivariant features encode chirality, and a self-supervised denoising objective enables pretraining on large unlabelled molecular datasets.

Our experiments demonstrate that simulation-trained representations transfer directly to quantitative structure-property prediction. Under a matched evaluation protocol on the TDC and MoleculeNet benchmarks, fine-tuned Rem3Di is the strongest model across the regression endpoints and competitive on most TDC classification tasks, with its advantage most pronounced on properties governed by 3D physics, such as solvation free energy. Ablations attribute this descriptor quality to denoising pretraining, whose benefit persists after parameter-efficient LoRA adaptation. The same descriptor extends naturally to chemistries where conventional 2D representations struggle: its pseudoscalar channels resolve stereochemistry that inversion-invariant features cannot, and it yields a chemically meaningful, interpretable organisation of transition-metal complexes without predefined bonding rules. Benchmarked on QM9-OR, a dataset of molecules with DFT-computed optical rotations, Rem3Di recovers the handedness of chiral centres with 77\% accuracy and the sign of the optical rotation with 70\% under a scaffold split, where two-dimensional fingerprint and descriptor baselines perform close to chance.

Predicting experimental properties remains difficult, limited by small and noisy labelled datasets. Denoising pretraining mitigates this scarcity and produces a descriptor space in which chemically similar molecules cluster together. Taken together, these results establish that the internal representations of atomistic foundation models, trained only to reproduce energies and forces, provide a transferable and chirality-aware foundation for molecular machine learning that reaches well beyond their original simulation setting.

\subsection*{Outlook} 

Smooth, differentiable molecular descriptors create a bridge between atomistic simulation and chemical machine learning. In the present work this bridge is tested on supervised property prediction and unsupervised chemical organisation. The same idea could support structure-aware screening, where a descriptor carries three-dimensional and stereochemical information into simple regression or classification models. Because Rem3Di remains differentiable with respect to atomic coordinates, its descriptors could also be evaluated as collective variables for enhanced sampling. Finally, the descriptor need not remain downstream of a frozen MLIP: future work could test joint training of the interatomic potential and descriptor, using self-supervised denoising as an auxiliary objective alongside energy and force regression. These extensions connect to work on training the interatomic potentials themselves with denoising objectives. As an auxiliary task, denoising generalised to non-equilibrium structures improves equivariant force fields~\cite{Liao2024}, and as a pretraining stage, denoising diffusion is already used to initialise production foundation potentials before energy and force training~\cite{Neumann2024, Rhodes2025}. Pretraining a descriptor and a potential jointly on unlabelled structures is therefore a natural next step. In general the Rem3Di framework points to a broader use for atomistic foundation-model representations, not only as inputs to molecular predictors, but as a reusable framework for learning, comparing, and navigating chemical structure.

\section*{Acknowledgements}

The authors thank Kieran Didi, Simon Mathis and Ilyes Batatia for useful discussion throughout the project.
S.W. kindly acknowledges support from the German Academic Scholarship Foundation, and travel sponsorship from Dunia Innovations. 
C.S. acknowledges support from the Isaac Newton Trust G122390, the UKRI reference EP/V062654/1 and the Royal Society RGS/R2/242614. 
L.L.S. kindly acknowledges support from Wolfson College, Cambridge through a Junior Research Fellowship and funding from the Advanced Research and Invention Agency (ARIA)
and the Department for Science, Innovation and Technology (DSIT) and Pillar VC under the Encode AI for Science Fellowship.

This work was also performed using resources provided by the Cambridge Service for Data Driven Discovery (CSD3) operated by the University of Cambridge Research Computing Service (www.csd3.cam.ac.uk), provided by funding from the Engineering and Physical Sciences Research Council (capital grant EP/T022159/1), and DiRAC funding from the Science and Technology Facilities Council (www.dirac.ac.uk).
Access to CSD3 was obtained through a University of Cambridge EPSRC Core Equipment Award EP/X034712/1

\section*{Code and data availability}
\label{sec:data-avail}
The code, including the pseudoscalar triple tensor-product construction, is available at \url{https://github.com/molsuit/Rem3Di}, with accompanying documentation at \url{https://molsuit.github.io/Rem3Di/}.

\section*{References}
\bibliographystyle{naturemag-doi}
\bibliography{remedi-arxiv}

\appendix
\onecolumngrid

\newpage
\section*{Appendix}

\section{Implementation details}
\label{app:impl}

\textbf{Online featurisation.} For each molecule the coordinates are centred and
placed in a non-periodic cubic box of side $2\,(\text{radius} + \max(20~\text{\AA},\, 1.5 \,r_\text{max}))$,
large enough that the MLIP neighbour list spans the whole molecule, per-system total
charge and spin multiplicity are passed to the potential. MLIP features are computed
online at every training step and detached. The online featurisation in principle allows for fine-tuning of the MACE features, or using the Rem3Di architecture to pretrain MLIPs.

\textbf{Numerical precision.} The MLIP runs in single precision in our
experiments. The equivariant tensor algebra of the chiral block benefits from extra
precision: the atomic preprocessor is cast to double precision in the
descriptor-inference and supervised-fine-tuning paths, while pretraining runs the
preprocessor in single precision. In all cases the per-atom embeddings are cast back
to single precision at the boundary that feeds the encoder.

\textbf{Normalisation statistics.} The invariant standardisation mean and
standard deviation are estimated once by streaming up to $\sim$$31$ batches of
$10{,}000$ structures, accumulating per-channel sums and sums-of-squares in double
precision, with the standard deviation floored at $10^{-9}$. The equivariant RMS
layer-norm is active only when the chiral branch is enabled.

\textbf{Warm-started geometry.} Both the pair-to-attention bias projection and the
gate of the pair feed-forward block are zero-initialised, so the encoder's first
forward pass is identical to a standard self-attention transformer and the geometric
pathway is learned from that starting point.

\section{Architecture and training hyperparameters}
\label{app:hyperparameters}

Table~\ref{tab:hparams} lists the architecture and training settings used in our
baseline configuration. Dimensions not listed are resolved automatically from the
MLIP irrep signature at model-assembly time.

\begin{table}[h]
\centering
\caption{Default Rem3Di architecture and training hyperparameters.}
\label{tab:hparams}
\footnotesize
\setlength{\tabcolsep}{4pt}
\renewcommand{\arraystretch}{1.05}
\begin{tabular}{ll}
\toprule
\multicolumn{2}{l}{\textit{Featurisation}} \\
\midrule
Foundation MLIP & frozen MACE (MACE-POLAR-1-M) \\
MLIP features used & deepest layer, $512{\times}0e + 512{\times}1o$ \\
Invariant projection & linear, $512\rightarrow256$ (bias-free) \\
\midrule
\multicolumn{2}{l}{\textit{Encoder}} \\
\midrule
Blocks $\times$ width & $4 \times 256$ \\
Attention heads & $8$ \\
Feed-forward width & $1024$ (GeGLU) \\
Dropout & $0.3$ \\
Pair width $d_\text{pair}$ & $64$ \\
Radial basis / cutoff & Bessel, $16$ functions, $32$~\AA{} \\
Chiral channels $K$ & $128$ (when enabled) \\
\midrule
\multicolumn{2}{l}{\textit{Pretraining}} \\
\midrule
Noise scale $\sigma$ & $0.3$ (constant) \\
Optimiser & AdamW, one-cycle schedule \\
Learning rate / weight decay & $3\times10^{-4}$ / $10^{-3}$ \\
Gradient-norm clip & $1.0$ \\
Epochs & $15$ \\
Batch cap / bucket size & $5250$ atoms / $512$ \\
VICReg ($\lambda_\text{var},\lambda_\text{cov}$, target std) & off; $25,\,1,\,1$ when on \\
\midrule
\multicolumn{2}{l}{\textit{Adaptation}} \\
\midrule
LoRA rank $r$ / scale $\alpha$ & $16$ / $32$ (attention + pool) \\
Classical-ML cross-validation & $5\times3$ outer, $3$-fold inner search \\
Chiral classifier loss & focal ($\gamma=2$, inverse-freq.\ weights) \\
\bottomrule
\end{tabular}
\end{table}

\section{Denoising pretraining algorithm}
\label{app:pretrain-pseudocode}

Algorithm~\ref{alg:pretrain} shows a single denoising pretraining step for Rem3Di. A frozen MLIP provides atom,
pair, and geometry features. The encoder produces a clean molecular descriptor,
and the decoder reconstructs clean atom features from noised atom features
conditioned on this descriptor.

\begin{center}
\begin{minipage}{0.7\linewidth}
\begin{algorithm}[H]
\DontPrintSemicolon
\SetAlgoNlRelativeSize{-1}
\SetKwComment{Comment}{$\triangleright$\ }{}
\caption{One denoising pretraining step for Rem3Di.}
\label{alg:pretrain}

\KwIn{
molecule batch $\mathcal{B}$;
frozen MLIP $\phi$;
trainable preprocessor $q$;
encoder $E_\theta$;
decoder $D_\psi$;
noise scale $\sigma$
}

\vspace{0.3em}

\KwOut{
updated $q$, $E_\theta$, and $D_\psi$;
frozen $\phi$ unchanged
}

\BlankLine
\textit{Frozen feature extraction.}

$(\mathbf{A},\,\mathbf{P},\,\mathbf{m})
\leftarrow \phi(\mathcal{B})$
\Comment*[r]{MLIP features and real-atom mask}

$\mathbf{S}^{0} \leftarrow q(\mathbf{A})$
\Comment*[r]{clean preprocessed atom features}

\BlankLine
\textit{Clean molecular descriptor.}

$\mathbf{M}
\leftarrow
E_\theta(\mathbf{S}^{0},\,\mathbf{P},\,\mathbf{m})$
\Comment*[r]{encode clean molecule}

\BlankLine
\textit{Feature-space corruption.}

$\boldsymbol{\epsilon} \sim \mathcal{N}(\mathbf{0}, \mathbf{I})$

$\widetilde{\mathbf{S}}^{0}
\leftarrow
\operatorname{sg}(\mathbf{S}^{0})
+
\sigma \boldsymbol{\epsilon}$
\Comment*[r]{noise valid atom features}

\BlankLine
\textit{Denoising decoder.}

$\widehat{\mathbf{S}}^{0}
\leftarrow
D_\psi\!\left(
\widetilde{\mathbf{S}}^{0},
\mathbf{M};
\operatorname{sg}(\mathbf{P}),
\mathbf{m}
\right)$
\Comment*[r]{reconstruct clean atom features}

\BlankLine
\textit{Training objective.}

$\mathcal{L}_{\mathrm{denoise}}
\leftarrow
\dfrac{1}{N_{\mathrm{atoms}}\sigma^{2}}
\sum_{i \in \mathrm{real}}
\left\lVert
\widehat{\mathbf{S}}^{0}_{i}
-
\mathbf{S}^{0}_{i}
\right\rVert^{2}$
\Comment*[r]{masked variance-normalised MSE}

$\mathcal{L}
\leftarrow
\mathcal{L}_{\mathrm{denoise}}$

\If{VICReg regularisation is used}{
$\mathcal{L}
\leftarrow
\mathcal{L}
+
\lambda_{\mathrm{var}}\mathcal{L}_{\mathrm{var}}(\mathbf{M})
+
\lambda_{\mathrm{cov}}\mathcal{L}_{\mathrm{cov}}(\mathbf{M})$
\Comment*[r]{discourage descriptor collapse}
}

\BlankLine
\textit{Parameter update.}

Update $q$, $\theta$, and $\psi$ using $\nabla \mathcal{L}$\;

Keep the MLIP $\phi$ frozen\;

\end{algorithm}
\end{minipage}
\end{center}

\newpage
\section{Stereochemistry tasks}

We train Rem3Di models on two stereochemical classification tasks. The first is the ChiralCat benchmark~\cite{Peng2025}, in which models classify the type of chirality among the classes achiral, central, axial, helical, and planar. The dataset contains structures with overlapping atoms, which required filtering and re-splitting. The helical and planar classes contain few labelled structures, so small changes in the model prediction alter the accuracy metrics substantially, and direct comparisons between Rem3Di and the baselines are of limited value. Rem3Di models with and without pseudoscalars both classify the type of chirality, which can be determined from atomic environments alone. The ChiralCat benchmark however does not distinguish between R and S configuration. For Rem3Di, the focal loss, which upweights the rare chirality classes, balances accuracy across the imbalanced classes. As the next benchmark shows, pseudoscalars are required for determining the handedness of the chirality.

\begin{table*}[h]
\centering
\caption{Per-class accuracy (\%) on the ChiralCat chirality-classification
benchmark. Literature methods (E3FP, Uni-Mol, ChiralCat) are taken from the
ChiralCat paper and evaluated on its test split. Rem3Di variants are our
ablation, evaluated on our own split, as the ChiralCat dataset contained defective molecular structures. \emph{Weighted} is the sample-weighted
(micro) accuracy over each method's own evaluation set. \emph{Bal.\ Acc.} is
the unweighted mean of the five per-class accuracies. Per-block support rows
give the number of test molecules per class. }
\label{tab:chiralcat}
\setlength{\tabcolsep}{6pt}
\renewcommand{\arraystretch}{1.15}
\begin{tabular}{l ccccc cc}
\toprule
& \multicolumn{5}{c}{Per-class accuracy} & \multicolumn{2}{c}{Overall} \\
\cmidrule(lr){2-6} \cmidrule(lr){7-8}
Method & Achiral & Central & Axial & Helical & Planar & Weighted & Bal.\ Acc. \\
\midrule
\addlinespace[2pt]
\textit{Support (ChiralCat test split)} & 140 & 139 & 107 & 8 & 12 & 406 & -- \\
\addlinespace[2pt]
E3FP + XGBoost & 90.00 & 87.05 & 63.55 & 0.00 & 50.00 & 79.06 & 58.12 \\
Uni-Mol        & 95.00 & 76.98 & 73.83 & 37.50 & 83.33 & 81.77 & 73.33 \\
ChiralCat      & 96.43 & 98.56 & 76.64 & 37.50 & 91.67 & 90.64 & 80.16 \\
\midrule
\addlinespace[2pt]
\textit{Support (our test split)} & 1000 & 625 & 55 & 4 & 6 & 1690 & -- \\
\addlinespace[2pt]
Rem3Di (ps\_focal)   & 95.60 & 96.30 & 92.70 & 50.00 & \textbf{100.00} & 95.67 & 86.93 \\
Rem3Di (nops\_focal) & 98.60 & 99.00 & \textbf{94.50} & 50.00 & \textbf{100.00} & 98.50 & 88.44 \\
Rem3Di (ps\_ce)      & 98.60 & 99.20 & 92.70 & 50.00 & \textbf{100.00} & 98.52 & 88.11 \\
Rem3Di (nops\_ce)    & \textbf{99.20} & \textbf{99.50} & \textbf{94.50} & \textbf{75.00} & 83.33 & \textbf{99.04} & \textbf{90.32} \\
\bottomrule
\end{tabular}
\end{table*}

\newpage

\section{Benchmarks}
\label{app:endpoints}

\subsection{ADMET and activity prediction tasks}

We evaluate Rem3Di on drug-property benchmarks from the Therapeutics Data Commons (TDC) ADMET suite~\cite{Huang2021} and MoleculeNet~\cite{Wu2018}. The TDC tasks cover absorption, distribution, metabolism, and toxicity endpoints relevant to small-molecule pharmacokinetics and safety. Binary endpoints are evaluated with AUROC, whereas regression endpoints are evaluated with MAE. The MoleculeNet regression tasks used here are ESOL and FreeSolv and are evaluated with RMSE.

\textbf{TDC classification endpoints.} \noindent{}
BBB denotes blood--brain-barrier penetration and is a binary measure of whether a compound crosses the blood--brain barrier. HIA denotes human intestinal absorption and measures whether a compound is absorbed after oral administration. Pgp denotes inhibition of P-glycoprotein, an efflux transporter involved in absorption, distribution, and multidrug resistance. Bioav. denotes oral bioavailability and measures whether a compound reaches systemic circulation after oral dosing. Tox-Avg is the mean AUROC over toxicity classification endpoints, including hERG blockade, Ames mutagenicity, and drug-induced liver injury. CYP-Avg is the mean AUROC over cytochrome P450 metabolism endpoints, covering CYP2C9, CYP2D6, and CYP3A4 inhibition and substrate classification. All classification targets are binary and dimensionless.

\textbf{TDC regression endpoints.}
LD50 is an acute oral toxicity endpoint, defined as the dose causing lethality in 50\% of the test population. The TDC label is reported as $\log(1/(\mathrm{mol}/\mathrm{kg}))$. Caco-2 measures apparent permeability across Caco-2 cell monolayers, an in vitro proxy for intestinal permeability. Its results are reported on the released TDC target scale. PPBR denotes plasma protein binding rate and is reported as a percentage. LIPO denotes lipophilicity, measured as the octanol/water distribution coefficient $\log D$ at pH 7.4, and is therefore dimensionless.

\textbf{MoleculeNet regression endpoints.}
ESOL measures aqueous solubility and is reported as $\log S$ in $\log(\mathrm{mol}/\mathrm{L})$. FreeSolv measures hydration free energy in water and is reported in $\mathrm{kcal}\,\mathrm{mol}^{-1}$. These two endpoints probe complementary physical properties: ESOL reflects bulk solubility, whereas FreeSolv is a direct thermodynamic measure of transferring a molecule from gas phase to water.

\subsection{Complete drug-property benchmark results}
\label{app:benchmark}

Table~\ref{tab:benchmark-full} reports the complete per-task TDC and MoleculeNet~\cite{Wu2018} leaderboard, including all external baselines, the endpoints summarised in the main text (Table~\ref{tab:benchmark}), and the MoleculeNet classification suite (BACE, BBBP, ClinTox, SIDER, Tox21) omitted there for compactness.

On these small MoleculeNet classification sets Rem3Di is competitive but does not lead: its frozen descriptor is the best entry on Tox21, whereas on BACE, BBBP and ClinTox it trails the strongly pretrained 2D graph models such as GROVER. Consistent with the limited training data on these endpoints, full fine-tuning \emph{degrades} small-classification performance relative to the frozen descriptor (e.g.\ ClinTox AUROC $0.86\!\rightarrow\!0.76$) while improving every regression endpoint, the contrast that motivates the adaptation study in the main text.

\begin{table*}[bth]
\centering
\caption{\textbf{Complete drug-property benchmark on TDC and MoleculeNet} (all baselines and endpoints). The main-text Table~\ref{tab:benchmark} shows only the TDC classification and the combined regression results. AUROC is used for classification ($\uparrow$), and MAE (TDC) / RMSE (MolNet) for regression ($\downarrow$). Cells: mean\dev{std} over seeds (our rows: 3 seeds, external baselines: as published). Each block is grouped into external published baselines (top) and matched-protocol evaluations (bottom). Within a block, the top-2 rankable cells per task are shaded \colorbox{firsthl}{first}/\colorbox{secondhl}{second}. $^{\dagger}$: coverage $<90\%$ (excluded from ranking). \textsc{Top2Cnt} counts a model's top-2 placements across that block's rankable cells.}
\label{tab:benchmark-full}
\scriptsize
\renewcommand{\arraystretch}{0.88}
\setlength{\tabcolsep}{3pt}
\begin{tabular}{l ccccccc}
\toprule
\multicolumn{8}{c}{\textsc{TDC classification, AUROC} $\uparrow$} \\
\midrule
\textsc{Model} & \textsc{BBB} & \textsc{HIA} & \textsc{Pgp} & \textsc{Bioav.} & \textsc{Tox-Avg} & \textsc{CYP-Avg} & \textsc{Top2Cnt/6} \\
\midrule
D-MPNN / Chemprop & 0.864\dev{.010} & \cellcolor{firsthl}0.976\dev{.004} & 0.889\dev{.005} & 0.617\dev{.050} & 0.821\dev{.019} & 0.819\dev{.004} & 1 \\
AttentiveFP & 0.855\dev{.011} & \cellcolor{secondhl}0.974\dev{.007} & 0.892\dev{.012} & \cellcolor{secondhl}0.632\dev{.039} & 0.842\dev{.010} & 0.749\dev{.008} & 2 \\
DeepMol & 0.774\dev{.023} & 0.880\dev{.012} & 0.821\dev{.007} & 0.509\dev{.026} & 0.735\dev{.015} & 0.770\dev{.008} & 0 \\
FPGNN & 0.888\dev{.018} & 0.958\dev{.012} & \cellcolor{firsthl}0.930\dev{.007} & \cellcolor{firsthl}0.666\dev{.035} & 0.860\dev{.017} & \cellcolor{secondhl}0.866\dev{.004} & 3 \\
TranFoxMol & 0.868\dev{.019} & 0.951\dev{.036} & 0.875\dev{.011} & 0.619\dev{.019} & 0.837\dev{.017} & 0.860\dev{.006} & 0 \\
\midrule
ECFP4 + LightGBM & 0.884\dev{.005} & 0.905\dev{.048} & 0.907\dev{.013} & 0.591\dev{.036} & 0.829\dev{.013} & \cellcolor{secondhl}0.866 & 1 \\
UniMol2 (frozen) & 0.884\dev{.014} & 0.902\dev{.088}$^{\dagger}$ & 0.898\dev{.017} & 0.617\dev{.041} & 0.833\dev{.020}$^{\dagger}$ & \cellcolor{firsthl}0.867 & 1 \\
MuMo (frozen) & 0.893\dev{.019} & 0.980\dev{.012}$^{\dagger}$ & 0.891\dev{.006} & 0.626\dev{.048} & 0.812\dev{.012}$^{\dagger}$ & 0.852 & 0 \\
\textbf{Rem3Di (frozen)} & \cellcolor{secondhl}0.894\dev{.004} & 0.949\dev{.019}$^{\dagger}$ & 0.898\dev{.012} & 0.558\dev{.057} & \cellcolor{secondhl}0.863\dev{.013} & 0.862 & 2 \\
\textbf{Rem3Di (fine-tuned)} & \cellcolor{firsthl}0.898\dev{.007} & 0.974\dev{.008}$^{\dagger}$ & \cellcolor{secondhl}0.911\dev{.006} & 0.561\dev{.018} & \cellcolor{firsthl}0.871\dev{.016} & \cellcolor{firsthl}0.867 & 4 \\
\midrule
\multicolumn{8}{c}{\textsc{MoleculeNet classification, AUROC} $\uparrow$} \\
\midrule
\textsc{Model} & \textsc{BACE-S} & \textsc{BBBP-S} & \textsc{ClinTox} & \textsc{SIDER} & \textsc{Tox21} & & \textsc{Top2Cnt/5} \\
\midrule
D-MPNN / Chemprop & 0.809\dev{.006} & 0.710\dev{.003} & 0.906\dev{.006} & 0.570\dev{.007} & 0.759\dev{.007} & & 0 \\
AttentiveFP & 0.784\dev{.022} & 0.643\dev{.018} & 0.847\dev{.003} & 0.606\dev{.032} & 0.761\dev{.005} & & 0 \\
GROVER & \cellcolor{firsthl}0.894\dev{.028} & \cellcolor{firsthl}0.940\dev{.019} & \cellcolor{firsthl}0.944\dev{.021} & 0.658\dev{.023} & 0.831\dev{.025} & & 3 \\
MolCLR & 0.824\dev{.009} & 0.722\dev{.021} & 0.912\dev{.035} & 0.589\dev{.014} & 0.750\dev{.002} & & 0 \\
GraphMVP & 0.812\dev{.009} & 0.724\dev{.016} & 0.791\dev{.028} & 0.639\dev{.012} & 0.759\dev{.005} & & 0 \\
GEM & 0.856\dev{.011} & 0.724\dev{.004} & 0.901\dev{.013} & \cellcolor{firsthl}0.672\dev{.004} & 0.781\dev{.001} & & 1 \\
Uni-Mol & \cellcolor{secondhl}0.857\dev{.002} & 0.729\dev{.006} & \cellcolor{secondhl}0.919\dev{.018} & 0.659\dev{.013} & 0.796\dev{.005} & & 2 \\
FPGNN & 0.831\dev{.011} & \cellcolor{secondhl}0.892\dev{.019} & 0.732\dev{.068} & \cellcolor{secondhl}0.661\dev{.014} & 0.833\dev{.004} & & 2 \\
TranFoxMol & 0.780\dev{.032} & 0.881\dev{.015} & 0.830\dev{.047} & 0.636\dev{.022} & 0.816\dev{.011} & & 0 \\
\midrule
ECFP4 + LightGBM & 0.795\dev{.043} & 0.794\dev{.119} & 0.818\dev{.074} & 0.647\dev{.025} & 0.805\dev{.015} & & 0 \\
UniMol2 (frozen) & 0.833\dev{.070}$^{\dagger}$ & 0.800\dev{.117} & 0.794\dev{.065}$^{\dagger}$ & 0.622\dev{.032} & 0.842\dev{.011} & & 0 \\
MuMo (frozen) & 0.787\dev{.019}$^{\dagger}$ & 0.661\dev{.008} & 0.800\dev{.075}$^{\dagger}$ & 0.599\dev{.013}$^{\dagger}$ & 0.827\dev{.012} & & 0 \\
\textbf{Rem3Di (frozen)} & 0.794\dev{.003} & 0.680\dev{.005} & 0.857\dev{.099} & 0.626\dev{.016}$^{\dagger}$ & \cellcolor{firsthl}0.861\dev{.014} & & 1 \\
\textbf{Rem3Di (fine-tuned)} & 0.795\dev{.002} & 0.657\dev{.008} & 0.763\dev{.054} & 0.616\dev{.027}$^{\dagger}$ & \cellcolor{secondhl}0.853\dev{.012} & & 1 \\
\midrule
\multicolumn{8}{c}{\textsc{TDC regression, MAE} $\downarrow$} \\
\midrule
\textsc{Model} & \textsc{LD50} & \textsc{Caco-2} & \textsc{PPBR} & \textsc{LIPO} & & & \textsc{Top2Cnt/4} \\
\midrule
D-MPNN / Chemprop & \cellcolor{secondhl}0.607\dev{.022} & 0.388\dev{.077} & \cellcolor{secondhl}8.158\dev{.314} & \cellcolor{firsthl}0.448\dev{.014} & & & 3 \\
AttentiveFP & 0.678\dev{.012} & 0.401\dev{.032} & 9.373\dev{.335} & 0.572\dev{.007} & & & 0 \\
DeepMol & \cellcolor{firsthl}0.589\dev{.006} & 0.327\dev{.012} & 9.533\dev{.162} & 0.660\dev{.004} & & & 1 \\
FPGNN & 0.638\dev{.024} & \cellcolor{secondhl}0.326\dev{.040} & 8.465\dev{1.709} & 0.544\dev{.011} & & & 1 \\
TranFoxMol & 0.645\dev{.036} & 0.487\dev{.068} & 9.055\dev{.523} & 0.525\dev{.024} & & & 0 \\
\midrule
ECFP4 + LightGBM & 0.659\dev{.005} & 0.477\dev{.011} & 9.646\dev{.122} & 0.627\dev{.014} & & & 0 \\
UniMol2 (frozen) & 0.682\dev{.012} & 0.338\dev{.018} & 8.985\dev{.244} & 0.602\dev{.017} & & & 0 \\
MuMo (frozen) & 0.693\dev{.030} & 0.353\dev{.022} & 8.265\dev{.287} & 0.677\dev{.008} & & & 0 \\
\textbf{Rem3Di (frozen)} & 0.656\dev{.015} & 0.368\dev{.022} & 8.536\dev{.313} & 0.625\dev{.006} & & & 0 \\
\textbf{Rem3Di (fine-tuned)} & 0.638\dev{.031} & \cellcolor{firsthl}0.326\dev{.016} & \cellcolor{firsthl}8.054\dev{.237} & \cellcolor{secondhl}0.499\dev{.015} & & & 3 \\
\midrule
\multicolumn{8}{c}{\textsc{MoleculeNet regression, RMSE} $\downarrow$} \\
\midrule
\textsc{Model} & \textsc{ESOL} & \textsc{FreeSolv} & & & & & \textsc{Top2Cnt/2} \\
\midrule
D-MPNN / Chemprop & 1.050\dev{.008} & 2.082\dev{.082} & & & & & 0 \\
AttentiveFP & 0.877\dev{.029} & 2.073\dev{.183} & & & & & 0 \\
GROVER & 0.831\dev{.120} & 1.544\dev{.397} & & & & & 0 \\
MolCLR & 1.271\dev{.040} & 2.594\dev{.249} & & & & & 0 \\
GraphMVP & 1.029\dev{.033} & -- & & & & & 0 \\
GEM & 0.798\dev{.029} & 1.877\dev{.094} & & & & & 0 \\
Uni-Mol & 0.788\dev{.029} & 1.480\dev{.048} & & & & & 0 \\
FPGNN & 0.658\dev{.006} & 1.106\dev{.195} & & & & & 0 \\
TranFoxMol & 0.930\dev{.261} & 1.225\dev{.155} & & & & & 0 \\
\midrule
ECFP4 + LightGBM & 1.069\dev{.124} & 2.178\dev{.400} & & & & & 0 \\
UniMol2 (frozen) & 0.655\dev{.065} & 1.169\dev{.144} & & & & & 0 \\
MuMo (frozen) & \cellcolor{secondhl}0.647\dev{.039} & 1.243\dev{.111} & & & & & 1 \\
\textbf{Rem3Di (frozen)} & 0.655\dev{.071} & \cellcolor{secondhl}1.069\dev{.205} & & & & & 1 \\
\textbf{Rem3Di (fine-tuned)} & \cellcolor{firsthl}0.594\dev{.057} & \cellcolor{firsthl}0.820\dev{.177} & & & & & 2 \\
\bottomrule
\end{tabular}
\end{table*}

\newpage

\end{document}